\documentclass[draftclsnofoot]{IEEEtran}
 \onecolumn
\usepackage{latexsym}
\usepackage{cite}
\usepackage{amssymb}
\usepackage{bm}
\usepackage{amsmath, amssymb}
\usepackage[dvips]{graphics}
\usepackage{graphicx}
\usepackage{psfrag}
\begin{document}
\title{Finite State Channels with Time-Invariant Deterministic Feedback}%We'll see
%\end{titlepage}
%\setlength{\textwidth}{7in} \setlength{\textheight}{9in}
%\setlength{\topmargin}{-0.4in} \setlength{\oddsidemargin}{-0.30in}
%\thispagestyle{empty} \setcounter{page}{1}
%\setlength{\baselineskip}{1\baselineskip} \maketitle
%\IEEEpeerreviewmaketitle
\author{Haim Permuter, Tsachy Weissman and Andrea Goldsmith\\
%Information Systems Laboratory\\
%Department of Electrical Engineering\\
%Stanford University\\
%Stanford, CA, 94305 USA \\
%Email: \{haim1, tsachy, andrea\}@stanford.edu
%\markboth{Journal
%of \LaTeX\ Class Files,~Vol.~1, No.~11,~November~2002}{Shell
%\MakeLowercase{\textit{et al.}}: Bare Demo of IEEEtran.cls for
%Journals}
\thanks{This work was partially
supported by NSF Grant CCR-0311633 and NSF CAREER grant.}
\thanks{The authors are with the Department of Electrical Engineering, Stanford University, Stanford, CA 94305, USA.
(Email: \{haim1, tsachy, andrea\}@stanford.edu)}
%\thanks{M. Shell is with the Georgia Institute of Technology.}}
}

\maketitle

\begin{abstract}
We consider  capacity of discrete-time channels with feedback for
the general case where the feedback is a time-invariant
deterministic function of the output samples. Under the assumption
that the channel states take values in a finite alphabet, we find
an achievable rate and an upper bound on the capacity. We further
show that when the channel is indecomposable, and has no
intersymbol interference (ISI), its capacity is given by the limit
of  the maximum of the (normalized) directed information between
the input $X^N$ and the output $Y^N$, i.e. $C = \lim_{N
\rightarrow \infty} \frac{1}{N} \max I(X^N \rightarrow Y^N )$,
where the maximization is taken over the causal conditioning
probability $Q(x^N||z^{N-1})$ defined in this paper. The capacity
result is used to show that the source-channel separation theorem
holds for time-invariant determinist feedback. We also show that
if the state of the channel is known both at the encoder and the
decoder then feedback does not increase capacity.
\end{abstract}

\begin{keywords}
Feedback capacity, directed information, causal conditioning,
code-tree, random coding, maximum likelihood, source-channel
coding separation.
\end{keywords}

\newtheorem{question}{Question}
\newtheorem{claim}{Claim}
\newtheorem{guess}{Conjecture}
\newtheorem{definition}{Definition}
\newtheorem{fact}{Fact}
\newtheorem{assumption}{Assumption}
\newtheorem{theorem}{Theorem}
\newtheorem{lemma}[theorem]{Lemma}
\newtheorem{ctheorem}{Corrected Theorem}
\newtheorem{corollary}[theorem]{Corollary}
\newtheorem{proposition}{Proposition}
\newtheorem{example}{Example}
\newtheorem{pfth}{Proof}

%\section{Directed Information}

\section{Introduction} \label{s_introduction}
Shannon showed in \cite{shannon56} that feedback does not increase
the capacity of a memoryless channel, and therefore the capacity
of a memoryless channel with feedback is given by maximizing the
mutual information between the input $X$, and the output $Y$, i.e.
$C=\max_{P(X)}I(X;Y)$. In the case that there is no feedback, and
the channel is an indecomposable Finite-State Channel (FSC), the
capacity was shown by Gallager \cite{Gallager68} and by Blackwell,
Breiman and Thomasian \cite{Blackwell58} to be
\begin{equation}\label{e_capacity_no_feedback}
C_{NF}=\lim_{N\rightarrow \infty} \frac{1}{N}
\max_{P(x^N)}I(X^N;Y^N).
\end{equation}
One might be tempted to think that for an FSC with feedback all
that changes for (\ref{e_capacity_no_feedback}) to characterize
capacity is the optimal input distribution, which now must depend
on the feedback.
%*However, the
%following simple example will show that even for a very simple
%channel, in fact the mutual information $I(X^N;Y^N)$ is
%inappropriate for characterizing capacity.
%%*It is perhaps natural to
%%assume that the capacity of a FSC with feedback should be of the
%form of maximum over all input distributions
%$\{P(x_i|x^{i-1},y^{i-1}), i=1\dots N\}$ of the mutual information
%$\frac{1}{N} I(X^N;Y^N)$.
However, the following simple counterexample shows that there are
cases in which the mutual information $I(X^N;Y^N)$ results in a
larger quantity than the capacity. Let us consider the case where
the channel has only one state which is a binary symmetric channel
(BSC) with probability of error 0.5. It is obvious that no
information can be transferred through this channel even with
feedback and, therefore, the capacity of the channel is zero.
However, it is easy to see that if we let the input to the channel
at time $i$ equal the output of the channel at time $i-1$, i.e.\
$X_i=Y_{i-1}$, for $i>2$, which is possible in the presence of
feedback, then it can be easily shown that
\begin{eqnarray}
\frac{1}{N} I(X^N;Y^N)    & = & \frac{1}{N} \sum _{i=1}^{N} \
 I(X_i;Y^N|X^{i-1}) %\geq  \frac{1}{N} \sum _{i=2}^{N} \ I(X_i;Y^N|X^{i-1})\\
 \nonumber\\
& = & \frac{1}{N}  \sum _{i=2}^{N} I(Y_{i-1};Y_{i-1}|Y^{i-2})
=\frac{N-1}{N}.
\end{eqnarray}
Therefore, we see that for this channel  $\lim_{N\to \infty} \max
\frac{1}{N} I(X^N;Y^N) = 1$, while the capacity of the channel is
zero. The reason such examples exist is that  $I(X^N;Y^N)$ is
measuring the mutual information between $X^N$ and $Y^N$,
including the mutual information that is due to the feedback and
not due to the channel. This example thus indicates that the
capacity of the channel with feedback must involve maximization
over an expression other then $I(X^N;Y^n)$.

In 1989 the directed information appeared in an implicit way in a
paper by Cover and Pombra \cite{Cover89}. In an intermediate step
\cite[eq. 52]{Cover89} they showed that the directed information
can be used to characterize the capacity of additive Gaussian
noise channels with feedback. However, the term directed
information was coined only a year later by Massey in a key paper
\cite{Massey90}.

In  \cite{Massey90}, Massey introduced directed information,
denoted by $I(X^N\rightarrow Y^N)$, which he attributes to Marko
\cite{Marko73}. Directed information, $I(X^N \rightarrow Y^N)$, is
defined as:
\begin{equation}\label{e_direct_def}
I(X^N \rightarrow Y^N) \triangleq  \sum_{i=1}^{N}
I(X^i;Y_i|Y^{i-1}).
\end{equation}
Massey showed that directed information is the same as mutual
information $I(X^N;Y^N)$ in the absence of feedback and it gives a
better upper bound on the information that the channel output
$Y^N$ gives about the source sequence in the presence of feedback.

Tatikonda, in his Ph.D. dissertation \cite{Tatikonda00},
generalized the capacity formula of Verd\'{u} and Han
\cite{Verdu94} that deals with arbitrary single-user channels
without feedback to the case of arbitrary single-user channels
with feedback by using the directed information formula.
%Tatikonda proved in his Ph.D. dissertation that the directed information rate is achievable by generalizing Feinstein's Lemma \cite{Fein54}.
Recently, the directed information formula was used by Yang,
Kav\u{c}i\'{c} and Tatikonda  \cite{Yang05} and by Chen and Berger
\cite{Chen05} to compute the feedback capacity for some special
finite-state channels.

Directed information also appeared recently in a rate distortion
problem. Following the competitive prediction of Weissman and
Merhav \cite{Weissman03}, Pradhan \cite{Pradhan04,Pradhan04b}
formulated a problem of source coding with feed-forward and showed
that directed information can be used to characterize the rate
distortion function for the case of feed-forward. Another source
coding context where directed information has arisen is the recent
work by Zamir et. al. \cite{zamir06}, which gives a linear
prediction representation for the rate distortion function of a
stationary Gaussian source

In this paper we extend the achievability proof given by Gallager
in \cite{Gallager68} for the case of a finite-state channel (FSC)
without feedback to the case of a FSC with feedback. We find an
upper bound on the error of the maximum likelihood decoder for a
FSC with time invariant deterministic feedback. We develop an
upper bound on the error which allows us to find an achievable
rate for the channel. In addition, we state an upper bound on the
capacity of the channel and show that when the state transition of
the FSC does not depend on the input,
%*the upper bound on the capacity equals to
the achievable rate equals the upper bound and hence equals the
channel capacity. The main contribution of our work is in showing
that the directed information, which was conjectured by Massey
\cite{Massey90} to be the capacity of a channel with feedback, is
achievable with a random coding scheme and maximum likelihood
decoding, for any time-invariant deterministic feedback.

Time-invariant feedback includes the cases of quantized feedback,
delayed feedback, and even noisy feedback where the noise is known
to the encoder. In addition, it allows a unified treatment of
capacity analysis for two ubiquitous cases: channels without
feedback and channels with perfect feedback. These two setting are
special cases of time-invariant feedback: in the first case the
time-invariant function of the feedback is the null function and
in the second case the time-invariant function of the feedback is
the identity function.

The capacity of some channels with channel state information at
the receiver and transmitter was derived by Caire and Shamai in
\cite{shamai99}. Note that if the channel state information can be
considered part of the channel output and fed beck to the
transmitter, then this case is a special case of a channel with
time invariant feedback.

The remainder of the paper is organized as follows. Section
\ref{s_preliminaries} defines the channel setting and the notation
throughout the paper. Section \ref{s_main_results} provides a
concise summary of the main results of the paper. Section
\ref{s_property_of_causal_cond} introduces several properties of
causal conditioning and directed
information\label{s_property_of_causal_cond} that are later used
in finding an achievable rate. Section
\ref{s_proof_of_achievability} provides the proof of achievability
of capacity of FSCs with time invariant feedback. Section
\ref{s_upper_bound_on_capacity} gives an upper bound on the
capacity. Section \ref{s_indeco} gives the capacity of an
indecomposable FSC without intersymbol interference (ISI). Section
\ref{s_feedbacz_and_side_information} considers the case of FSCs
with feedback and side information and shows that if the state is
known both at the encoder and decoder then feedback does not
increase the capacity of the channel. Section
\ref{s_source_channel_sep} shows that optimality of source-channel
separation holds in the presence of time-invariant feedback. We
conclude in Section \ref{s_conclusion} with a summary of this work
and some related future directions.

\begin{figure}{
 \psfrag{v1\r}[][][0.8]{$m$}\psfrag{w1\r}[][][0.8]{Message}
 \psfrag{u1\r}[][][0.8]{Encoder}
\psfrag{d1\r}[][][0.8]{$x_i(m,z^{i-1})$}
\psfrag{v2\r}[][][0.8]{$x_i$} \psfrag{w2\r}[][][0.8]{$$}
 \psfrag{u2\r}[][][0.8]{Finite State Channel}
\psfrag{d2\r}[][][0.8]{$P(y_i,s_i|x_i,s_{i-1})$}
\psfrag{v3\r}[][][0.8]{$y_i$}
\psfrag{w3\r}[][][0.8]{$$}\psfrag{v1\r}[][][0.8]{$m$}
\psfrag{u3\r}[][][0.8]{Decoder} \psfrag{d3\r}[][][0.8]{$\hat
m(y^N)$} \psfrag{u5\r}[][][0.8]{Feedback Generator}
\psfrag{v5\r}[][][0.8]{$z_i(y_{i})$} \psfrag{v4\r}[][][0.8]{$y_i$}
\psfrag{w4\r}[][][0.8]{$$} \psfrag{u4\r}[][][0.8]{Unit Delay}
\psfrag{D5\r}[][][0.8]{Time-Invariant}
\psfrag{D6\r}[][][0.8]{Function}
\psfrag{v7\r}[][][0.8]{$z_{i-1}(y_{i-1})$}
 \psfrag{w5\r}[][][0.8]{$$}
\psfrag{v6\r}[][][0.8]{$\hat m$}\psfrag{w6a\r}[][][0.8]{Estimated}
\psfrag{w6b\r}[][][0.8]{Message} \centering
\includegraphics[width=16cm]{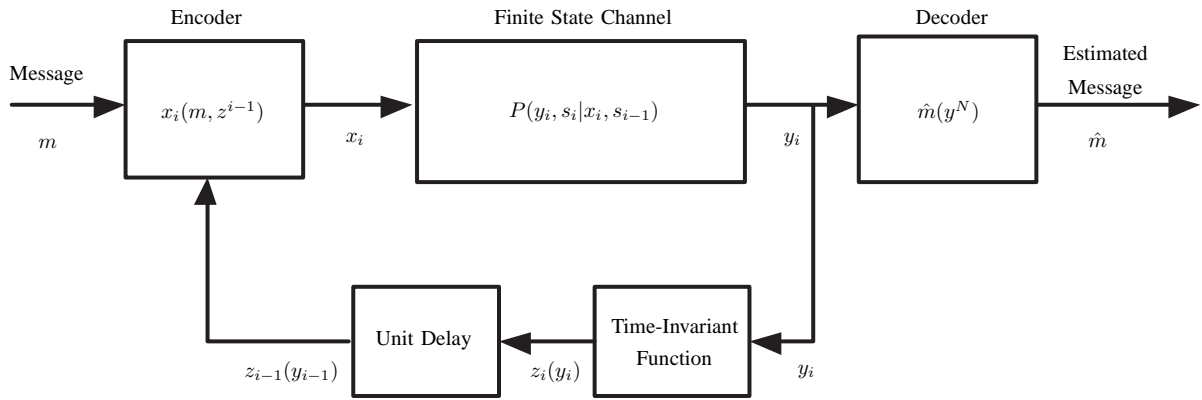}
\centering \caption{Channel with feedback that is a time invariant
deterministic function of the output. } \label{f_1}
}\end{figure} %%when I save on visio I used AI 3 - and I had to restart the computer and not all visio features work with AI3

\section{Channel Models and Preliminaries}\label{s_preliminaries} We use subscripts and
superscripts to denote vectors in the following way:
$x^i=(x_1\dots x_i)$ and $x_i^j=(x_i\dots x_j) $ for $i\leq j$.
For $i\leq0$, $x^i$ defines the null string as does $x_i^j$ when
$i>j$. Moreover, we use lower case to denote sample values and
upper case to denote random variables. Probability mass functions
are denoted by $P$ or $Q$ when the arguments specify the
distribution, e.g. $P(x|y)=P(X=x|Y=y)$. In this paper, we consider
only FSCs. The FSCs are a class of channels rich enough to include
channels with memory, e.g. channels with intersymbol interference.
The input of the channel is denoted by $\{X_1,X_2, \dots\}$, and
the output of the channel is denoted by $\{Y_1,Y_2, \dots\}$, both
taking values in a finite alphabet. In addition, the channel
states take values in a finite set of possible states.
 The channel is stationary and is characterized by a conditional probability assignment
 $P(y_i,s_i|x_i,s_{i-1})$ that satisfies
\begin{equation}
P(y_i,s_i|x^i,s^{i-1},y^{i-1})=P(y_i,s_i|x_i,s_{i-1}).
\end{equation}
An FSC is said to be without intersymbol interference (ISI) if the
input sequence does not affect the evolution of the state
sequence, i.e.\ $P(s_i|s_{i-1},x_i)=P(s_i|s_{i-1})$.

 We assume a
communication setting that includes feedback as shown in Fig.
\ref{f_1}. The transmitter (encoder) knows at time $i$ the message
$m$ and the feedback samples $z^{i-1}$. The output of the encoder
at time $i$ is denoted by $x_i$ and it is a function of the
message and the feedback. The channel is an FSC and the output of
the channel $y_i$ enters the decoder (receiver). The feedback
$z_i$ is a known time-invariant deterministic function of the
current output of the channel $y_i$. For example, $z_i$ could
equal $y_i$ or a quantized version of it. The encoder receives the
feedback sample with one unit delay.
%Later in the paper (Section
%\ref{s_tuple}) we also consider the case where the feedback is a
%deterministic function of more than one output sample i.e.
%$z_i(y_{i-M+1}^i)$ where $M$ is a constant.

Throughout this paper we use the {\it Causal Conditioning}
notation $(\cdot || \cdot)$, which was introduced and employed by
Kramer \cite{Kramer03,Kramer88} and by Massey \cite{Massey05}:
\begin{equation} \label{e_causal_cond_def}
P(y^N||x^N)\triangleq \prod_{i=1}^{N} P(y_i|x^{i},y^{i-1}).
\end{equation}
In addition, we introduce the following notation:
\begin{equation} \label{e_causal_cond_delay_def}
P(y^N||x^{N-1})\triangleq \prod_{i=1}^{N} P(y_i|x^{i-1},y^{i-1}).
\end{equation}
The definition given in (\ref{e_causal_cond_delay_def}) can be
considered to be a particular case of the definition given in
(\ref{e_causal_cond_def}) where $x_0$ is set to a dummy zero. This
concept was captured by a notation of Massey in \cite{Massey05}
via a concatenation at the beginning of the sequence $x^{N-1}$
with a dummy zero.
%* Where $y_0$ A more accurate notation inspired by
%\cite{Massey05} of $P(y^N||x^{N-1})$ would be $P(y^N||0 *
%x^{N-1})$, where $*$ denotes concatenation. This notation is more
%accurate because it emphasizes that the first element $x_1$ is
%aligned with the "dummy" zero.
The directed information $I(X^N\to Y^N)$ is defined in
(\ref{e_direct_def}) and, by using the definitions, we can express
directed information in terms of causal conditioning as
\begin{equation}\label{e_directed_def}
I(X^N \rightarrow Y^N)=\sum_{i=1}^{N} I(X^i;Y_i|Y^{i-1}) = \mathbf
E\left[ \log \frac{P(Y^N||X^N)}{P(Y^N)} \right],
\end{equation}
where $\mathbf E$ denotes expectation.% The directed information
%between $X^N$ and $Y^N$, causally conditioned on $Z^N$,
% is denoted as $I(X^N \to Y^N||Z^N)$ and  defined as:
%\begin{equation}
%I(X^N \to Y^N||Z^N)\triangleq \sum_{i=1}^{N}
%I(Y_i;X^i|Y^{i-1},Z^i).
%\end{equation}
The directed information between $X^N$ and $Y^N$, conditioned on
$S$, is denoted as $I(X^N \to Y^N|S)$ and is defined as:
\begin{equation}
I(X^N \to Y^N|S)\triangleq \sum_{i=1}^{N} I(Y_i;X^i|Y^{i-1},S).
\end{equation}

\section{Main Results} \label{s_main_results} In this section, we state the main results
of the paper.
\begin{itemize}
\item {\bf Causal conditioning and directed information:} In
Section \ref{s_property_of_causal_cond} we establish some
properties of causal conditioning and directed information that
are used throughout the proofs, and also provide some intuition
about the meaning of these terms.

\item {\bf Achievable rate:} For any finite-state channel with an initial state denoted by $s_0$, and with the feedback setting of
Fig. \ref{f_1}, and for any $R, 0\leq R < \underline C $, where
$\underline C$ is given by
 \begin{equation}
 \underline C = \lim_{N \rightarrow \infty} \frac{1}{N}  \max_{Q(x^N||z^{N-1})} \min_{s_0}I(X^N
\rightarrow Y^N |s_0)
 \end{equation}
(a limit that can be shown to exist), and any $\epsilon > 0$,
there exists an $(N,M)$
 block code such that for all messages $m,1\leq m \leq M=\lfloor 2^{NR} \rfloor$ and all initial
 states,
 the decoding error is upper bounded by $\epsilon$. This achievability result is establish via analysis of a random coding scheme with maximum likelihood
 decoding.
 % similar Gallager's analyze for FSC without feedback \cite{Gallager68}.
 %The proofs and the result are similar to Gallager's for FSC without feedback
% \cite{Gallager68} except substitute directed information (DI) for mutual information and make appropriate change to input distribution optimization.

 %of
% the achievable rate can be shown under a random coding scheme and maximum likelihood
% decoder.
% %, similar to the analyze given by Gallager
 %\cite{Gallager68}.

\item {\bf Converse:} For any given channel with the feedback as
in Fig. \ref{f_1}, any sequence of $(N,2^{NR})$ codes with
probability of decoding error that goes to zero as $N\to \infty$
must have
\begin{equation}
 R\leq  \lim_{N \rightarrow \infty} \frac{1}{N}  \max_{Q(x^N||z^{N-1})} I(X^N
\rightarrow Y^N )
 \end{equation}
where the limit is shown to exist.

\item {\bf Capacity:} For an indecomposable FSC without ISI, the achievable rate and the upper
bound are equal. Hence the capacity, which is defined as the
supremum of all achievable rates of the channel, is given by:
\begin{equation} \label{e_capacity_directed}
  C = \lim_{N \rightarrow \infty} \frac{1}{N}  \max_{Q(x^N||z^{N-1})} I(X^N
\rightarrow Y^N ).
 \end{equation}

\item {\bf State information and feedback:} Feedback does not increase the capacity of a strongly connected
FSC (every state can be reached from every other state with
positive probability under some input distribution) when the state
of the channel is known both at the encoder and the decoder.

\item{\bf Source-channel separation} Source-channel coding
separation is optimal for any channel with time-invariant
deterministic feedback where the capacity is given by eq.
(\ref{e_capacity_directed}).
\end{itemize}

\section{Properties of causal conditioning and directed information\label{s_property_of_causal_cond}} In this section we present some
properties of the causal conditioning distribution and the
directed information which are defined in Section
\ref{s_preliminaries} in eq. (\ref{e_causal_cond_def}),
(\ref{e_causal_cond_delay_def}) and (\ref{e_directed_def}). The
properties are used throughout the proof of achievability and also
help in gaining some intuition about those definitions and their
role in the proof of the achievability.

\begin{lemma}\label{l_QP}{\it Chain rule for causal conditioning.}
For any random variables $(X^N,Y^N)$
\begin{equation}
P(x^N,y^N)=P(y^N||x^N)P(x^N||y^{N-1}),
\end{equation}
and, consequently, if $Z^N$ is a random vector that satisfies
$P(x^N||y^{N-1})=P(x^N||z^{N-1})$ then
\begin{equation}
P(x^N,y^N)=P(y^N||x^N)P(x^N||z^{N-1}).
\end{equation}

\end{lemma}\label{l_1}
\begin{proof}
\begin{eqnarray}
P(y^N,x^N)& = &\prod_{i=1}^{N}P(y_i,x_i|x^{i-1},y^{i-1}) \nonumber \\
& = &\prod_{i=1}^{N}P(y_i|x^i,y^{i-1})P(x_i|x^{i-1},y^{i-1})\nonumber\\
%& = &\prod_{i=1}^{N}P(y_i|x^i,y^{i-1})P(x_i|x^{i-1},y^{i-1},z^{i-1}(y^{i-1}))\nonumber\\
%&\stackrel{(a)}{=}&\prod_{i=1}^{N}P(y_i|x^i,y^{i-1})P(x_i|x^{i-1},z^{i-1})\nonumber\\
&=&P(y^N||x^N)P(x^N||y^{N-1}).
\end{eqnarray}
%Equality (a) is due to the assumption of the lemma.
\end{proof}
%\begin{lemma}\label{l_QP}{\it Chain rule for causal conditioning.}
%For any random variables $(X^N,Y^N,Z^{N-1})$ that satisfies
%$P(x^N||y^{N-1})=P(x^N||z^{N-1})$,
%\begin{equation}
%P(x^N,y^N)=P(y^N||x^N)P(x^N||z^{N-1}),
%\end{equation}
%and if choose $z^N=y^N$ then we get that for any random variables
%$(X^N,Y^N)$
%\begin{equation}
%P(x^N,y^N)=P(y^N||x^N)P(x^N||y^{N-1}).
%\end{equation}
%
%\end{lemma}\label{l_1}
%\begin{proof}
%\begin{eqnarray}
%P(y^N,x^N)& = &\prod_{i=1}^{N}P(y_i,x_i|x^{i-1},y^{i-1}) \nonumber \\
%& = &\prod_{i=1}^{N}P(y_i|x^i,y^{i-1})P(x_i|x^{i-1},y^{i-1})\nonumber\\
%%& = &\prod_{i=1}^{N}P(y_i|x^i,y^{i-1})P(x_i|x^{i-1},y^{i-1},z^{i-1}(y^{i-1}))\nonumber\\
%&\stackrel{(a)}{=}&\prod_{i=1}^{N}P(y_i|x^i,y^{i-1})P(x_i|x^{i-1},z^{i-1})\nonumber\\
%&=&P(y^N||x^N)P(x^N||z^{N-1})
%\end{eqnarray}
%Equality (a) is due to the assumption of the lemma.
%\end{proof}
Note that there exists an analogy between this lemma and the chain
rule $P(x^N,y^N)=P(y^N|x^N)P(x^N)$. The analogy between the term
$P(y^N|x^N)$ and the term $P(y^N||x^N)$, and  between the term
$P(x^N)$ and the term $P(x^N||y^{N-1})$, can be helpful for
deriving equalities for the case of causal conditioning
distributions that are analogous to the equalities that hold for
regular distributions.

Let us define,
\begin{equation} \label{e_def1}
P(y^N||x^N,s)\triangleq \prod_{i=1}^{N} P(y_i|x^{i},y^{i-1},s).
\end{equation}

\begin{lemma}\label{l_QP_s0}
For any random variables $(X^N,Y^N,Z^{N-1},S_0)$ that satisfy
$P(x^N||y^{N-1},s_0)=P(x^N||z^{N-1})$,
\begin{equation}
P(x^N,y^N|s_0)=P(y^N||x^N,s_0)P(x^N||z^{N-1})
\end{equation}
\end{lemma}\label{l_1}
The proof of Lemma \ref{l_QP_s0} is similar to that of Lemma
\ref{l_QP} and therefore is omitted.
%\begin{proof}
%\begin{eqnarray}
%P(y^N,x^N|s_0)& = &\prod_{i=1}^{N}P(y_i,x_i|x^{i-1},y^{i-1},s_0) \nonumber \\
%& = &\prod_{i=1}^{N}P(y_i|x^i,y^{i-1},s_0)P(x_i|x^{i-1},y^{i-1},s_0)\nonumber\\
%& = &\prod_{i=1}^{N}P(y_i|x^i,y^{i-1},s_0)P(x_i|x^{i-1},y^{i-1},z^{i-1}(y^{i-1}),s_0)\nonumber\\
%&\stackrel{(a)}{=}&\prod_{i=1}^{N}P(y_i|x^i,y^{i-1},s_0)P(x_i|x^{i-1},z^{i-1}(y^{i-1}))\nonumber\\
%&=&P(y^N||x^N,s_0)Q(x^N||z^{N-1})
%\end{eqnarray}
%Equality (a) is due to the assumption on the probability
%assignment given in eq. \ref{e_as} and the assumption that the
%assigning probability of $x$ does not depend on the initial state
%$s_0$.
%\end{proof}

\begin{lemma}\label{l_sumQ}
{\it Causal conditioning is in the unit simplex.} For any random
variables $(X^N,Z^{N-1})$,
\begin{equation}
\sum_{x^N}P(x^N||z^{N-1})=1
\end{equation}
\end{lemma}
\begin{proof}
\begin{eqnarray}\label{e_Qsum}
\sum_{x^N}P(x^N||z^{N-1})& = & \sum_{x_1}\sum_{x_2}\dots\sum_{x_N}
\prod_{i=1}^N P(x_i|x^{i-1},z^{i-1}) \nonumber \\
& = & \sum_{x_1}\sum_{x_2}\dots\sum_{x_{N-1}} \left[
\left(\prod_{i=1}^{N-1} P(x_i|x^{i-1},z^{i-1})\right)\cdot \sum_{x_N} P(x_N|x^{N-1},z^{N-1})\right] \nonumber \\
& = & \sum_{x_1}\sum_{x_2}\dots\sum_{x_{N-1}} \
\left(\prod_{i=1}^{N-1} P(x_i|x^{i-1},z^{i-1})\right)\cdot 1 \nonumber \\
& = & \sum_{x^{N-1}}P(x^{N-1}||z^{N-2}).
\end{eqnarray}
In addition, $\sum_{x_1}P(x_1)=1$. Hence, by induction,
$\sum_{x^N}P(x^N||z^{N-1})=1$.
\end{proof}

\begin{lemma}
There is a one to one correspondence between causal conditioning
$P(x^N||z^{N-1})$ and the sequence of conditional distributions
$\{P(x_i|x^{i-1},z^{i-1})\}_{i=1}^{N}$.
\end{lemma}
\begin{proof}
It is obvious that the sequence
$\{P(x_i|x^{i-1},z^{i-1})\}_{i=1}^{N}$ determines the term
$P(x^N||z^{N-1})$. In the other direction we can use the proof of
Lemma \ref{l_sumQ}, in which we showed that $P(x^{N-1}||z^{N-2})$
is uniquely determined from $P(x^N||z^{N-1})$ by a summation over
$x_N$. Furthermore, by induction it can be shown that the sequence
$\{P(x^i||z^{i-1})\}_{i=1}^{N}$ is uniquely derived from
$P(x^N||z^{N-1})$ and then we can use the equality
\begin{equation}
P(x_i|x^{i-1},z^{i-1})=\frac{P(x^i||z^{i-1})}{P(x^{i+1}||z^{i})}
\end{equation}
to derive uniquely the sequence
$\{P(x_i|x^{i-1},z^{i-1})\}_{i=1}^{N}$.
\end{proof}
This lemma shows that the maximization in the capacity expressions
can be done on the set of sequences
$\{P(x_i|x^{i-1},z^{i-1})\}_{i=1}^{N}$ or, equivalently, on the
set of terms $P(x^N||z^{N-1})$. The lemma is analogous to the fact
that maximization over the set $P(x^N)$ is equivalent to
maximization over the set of sequences
$\{P(x_i|x^{i-1})\}_{i=1}^{N}$

\begin{lemma}\label{l_dirB}
Let $X^N,Y^N,Z^N$ be arbitrary random vectors and $S$ a random
variable taking values in an alphabet of size $|\mathcal{S}|$.
Then

\begin{equation}
\left|I(X^N \rightarrow Y^N||Z^{N-1})- I(X^N \rightarrow
Y^N||Z^{N-1},S)\right|\leq H(S)\leq
 \log|\mathcal{S}|.
\end{equation}
In particular, if $Z^N$ is $Y^N$, we get
\begin{equation}
\left|I(X^N \rightarrow Y^N)- I(X^N \rightarrow Y^N|S)\right|\leq
H(S)\leq \log|\mathcal{S}|.
\end{equation}
\end{lemma}

This lemma has an important role in the proofs for the capacity of
FSCs, because it bounds by a constant the difference of directed
information before and after conditioning on a state. The proof of
the lemma is given in Appendix \ref{a_l_dirB}.

The proof of the achievable rate for a channel with time-invariant
feedback $z_i(y_i)$ is an extension of the proof of the achievable
rate for a channel without feedback given in
\cite[Ch.5]{Gallager68}. Roughly speaking, in each step we have to
justify replacement of $Q(x^N)$ by $Q(x^N||z^{N-1})$ and of
$P(y^N|x^N)$ by $P(y^N||x^N)$. The replacement does not work in
all cases, for instance it does not work in the case of Theorem
4.6.4 in \cite{Gallager68}. At the end of the proof we will see
that the achievable rate is the same expression as the mutual
information with the probability mass function $Q(x^N)$ replaced
by $Q(x^N||z^{N-1})$ and $P(y^N|x^N)$ replaced by $P(y^N||x^N)$.
The following lemma shows that the
replacement results in directed information.%the directed information is the same expression as mutual
%information when the probability mass functions $Q(x^N)$ and
%$P(y^N|x^N)$ are replaced by $Q(x^N||z^{N-1})$ and $P(y^N||x^N)$,
%respectively.

\begin{lemma}\label{l_dir_mut}
Denote:
 \begin{equation}
\mathcal{I}(Q(x^{N}||z^{N-1}),P(y^N||x^N)) \triangleq  \sum_{y^N}
\sum_{x^N} Q(x^N||z^{N-1}) P(y^N||x^N)\log \frac{
P(y^N||x^N)}{\sum_{x^N} Q(x^N||z^{N-1}) P(y^N||x^N)},
\end{equation}
if $P(x^{N}||y^{N-1})=Q(x^{N}||z^{N-1})$ then, \begin{equation}
\mathcal{I}(Q(x^{N}||z^{N-1}),P(y^N||x^N))=I(X^N \rightarrow Y^N),
\end{equation}
 and similarly,
 \begin{equation}
\mathcal{I}(Q(x^{N}||z^{N-1}),P(y^N||x^N,s_0))=I(X^N \rightarrow
Y^N|s_0)
\end{equation}
\end{lemma}

\begin{proof}
\begin{eqnarray}\label{e_mut}
\mathcal{I}(Q(x^{N}||z^{N-1}),P(y^N||x^N)) &\triangleq
& \sum_{y^N} \sum_{x^N} Q(x^N||z^{N-1}) P(y^N||x^N)\log \frac{ P(y^N||x^N)}{\sum_{x^N} Q(x^N||z^{N-1}) P(y^N||x^N)}\nonumber \\
& \stackrel{(a)}{=}&  \mathbf E\left[\log \frac{ P(Y^N||X^N)}{\sum_{x^N} Q(x^N||z^{N-1}) P(Y^N||x^N)}\right] \nonumber\\
& \stackrel{}{=}& \mathbf E\left[\log P(Y^N||X^N)\right]-\mathbf
E\left[\log{\sum_{x^N}
Q(x^N||z^{N-1}) P(Y^N||x^N)}\right] \nonumber \\
& \stackrel{(b)}{=}&  \mathbf E \left[\log
\prod_{i=1}^{N} P(Y_i|X^{i},Y^{i-1})\right]-\mathbf E\left[\log P(Y^N) \right] \nonumber \\
&=&  \sum_{i=1}^{N}  \large[ \mathbf E [\log
P(Y_i|X^{i},Y^{i-1})]-
\mathbf E[ \log P(Y_i|Y^{i-1})] \large]  \nonumber \\
& =&  \sum_{i=1}^{N}
I(X^i;Y_i|Y^{i-1}) \nonumber \\
& = & I(X^N \rightarrow Y^N),
\end{eqnarray}
equalities (a) and (b) are due to Lemma \ref{l_QP}.
\end{proof}
The following lemma is an extension of the conservation law of
information given by Massey in \cite{Massey05}.
\begin{lemma}\label{l_conservation}
{\it Extended conservation law.} For any random variables
$(X^N,Y^N,Z^{N-1})$ that satisfy
$P(x^N||y^{N-1})=P(x^N||z^{N-1})$,
\begin{equation}
I(X^N;Y^N)=I(X^N \to Y^N)+I(\{0,Z^{N-1}\} \to X^N),
\end{equation}
where $\{0,Z^{N-1}\}$ is a concatenation of dummy zero to the
beginning of the sequence $Z^{N-1}$.
\end{lemma}
\begin{proof}
\begin{eqnarray}
I(X^N;Y^N)&\stackrel{(a)}{=}& \mathbf E\left[ \log \frac{
P(Y^N,X^N)}{ P(Y^N)P(X^N)}
\right]\nonumber\\
&\stackrel{(b)}{=}& \mathbf E\left[ \log \frac{
P(Y^N||X^N)P(X^N||Z^{N-1})}{ P(Y^N)P(X^N)}
\right]\nonumber\\
&\stackrel{}{=}& \mathbf E\left[ \log \frac{ P(Y^N||X^N)}{
P(Y^N)}\right]+\mathbf E\left[ \log \frac{ P(X^N||Z^{N-1})}{
P(X^N)}
\right]\nonumber\\
&\stackrel{(c)}{=}& I(X^N \to Y^N)+I(\{0,Z^{N-1}\} \to X^N).
\end{eqnarray}
Equality (a) is due to the definition of mutual information.
Equality (b) is due to Lemma \ref{l_QP}, and equality (c) is due
to the definition of directed information.
\end{proof}

The lemma was proven by induction in \cite{Massey05} for the case
where $z_i=y_i$. Here we see that the conservation law holds more
generally when $P(x^n||y^{n-1})=P(x^n||z^{n-1})$. In Subsection
\ref{sub_random_generation} we argue that this equality holds for
the setting of deterministic feedback $z_i(y_i)$ and therefore the
conservation law holds for the communication setting given in Fig
\ref{f_1}. This lemma is not used for the proof of achievability,
however, it gives a nice intuition for the relation of directed
information and mutual information in the setting of deterministic
feedback. In particular, the lemma implies that the mutual
information between the input and the output of the channel is
equal to the sum of directed information in the forward link and
the directed information in the backward link. In addition, it is
straightforward to see that in the case of no feedback, i.e. when
$z_i$ is null, then $I(X^N;Y^N)=I(X^N \to Y^N)$.

\section{Proof of Achievability \label{s_proof_of_achievability}}
The proof of the achievable rate of a channel with feedback given
here is an extension of the upper bound on the error of maximum
likelihood decoding derived by Gallager in \cite[Ch.5]{Gallager68}
for FSCs without feedback to the case of FSCs with feedback. The
main difference is that for analyzing the new coding scheme, the
feedback $z^{i-1}$ must be taken into account.

Let us first present a short outline of the proof:
\begin{itemize}
\item {\it Encoding scheme.} We randomly generate an encoding scheme for blocks of length $N$ by using the causal conditioning distribution
$Q(x^N||z^{N-1})$.
\item {\it Decoding.} We assume a maximum likelihood decoder and we denote the
error probability when message $m$ is sent and the initial state
of the channel is $s_0$ as $P_{e,m}(s_0)$.
\item{\it Bounding the error probability.} We show that for each $N>N(\epsilon)$, there exists a code for which we can bound the error
probability for all messages $1\leq m \leq \lfloor 2^{NR}\rfloor$
and all initial states $s_0$ by the following exponential,
\begin{equation}
P_{e,m}(s_0)\leq 2^{-N[E_r(R)-\epsilon]}.
\end{equation}
In addition, we show that if $R< \underline C$ then $E_r(R)$ is
strictly positive and, hence, by choosing $\epsilon<{E_r(R)}$, the
probability of error diminishes exponentially for $N>N(\epsilon)$.
\end{itemize}

\subsection{Random generation of coding
scheme}\label{sub_random_generation} 
\begin{figure}[h]{
\psfrag{c1}[][][1]{$x_1=0$} \psfrag{a1}[][][1]{$$}
\psfrag{c2}[][][1]{$x_2=1$} \psfrag{a2}[][][1]{$i=1$}
\psfrag{c3}[][][1]{$x_3=1$} \psfrag{a3}[][][1]{$i=2$}
\psfrag{c4}[][][1]{$x_4=0$} \psfrag{a4}[][][1]{$i=3$}

\psfrag{d1}[][][1]{$x_1=0$} \psfrag{a1}[][][1]{$$}
\psfrag{d2}[][][1]{$x_2=1$} \psfrag{a1}[][][1]{$$}
\psfrag{d3}[][][1]{$x_2=1$} \psfrag{a1}[][][1]{$$}
\psfrag{d4}[][][1]{$x_3=0$} \psfrag{a1}[][][1]{$$}
\psfrag{d5}[][][1]{$x_3=1$} \psfrag{a1}[][][1]{$$}
\psfrag{d6}[][][1]{$x_3=1$} \psfrag{a1}[][][1]{$$}
\psfrag{d7}[][][1]{$x_3=1$} \psfrag{a1}[][][1]{$$}
\psfrag{k0}[][][0.8]{$z_{i-1}=0$}
\psfrag{k1}[][][0.8]{$z_{i-1}=1$}
\psfrag{codeword}[][][1]{Codeword (case of no feedback)}
\psfrag{code-tree}[][][1]{Code-tree (case of feedback)}

\centerline{\includegraphics[width=16cm]{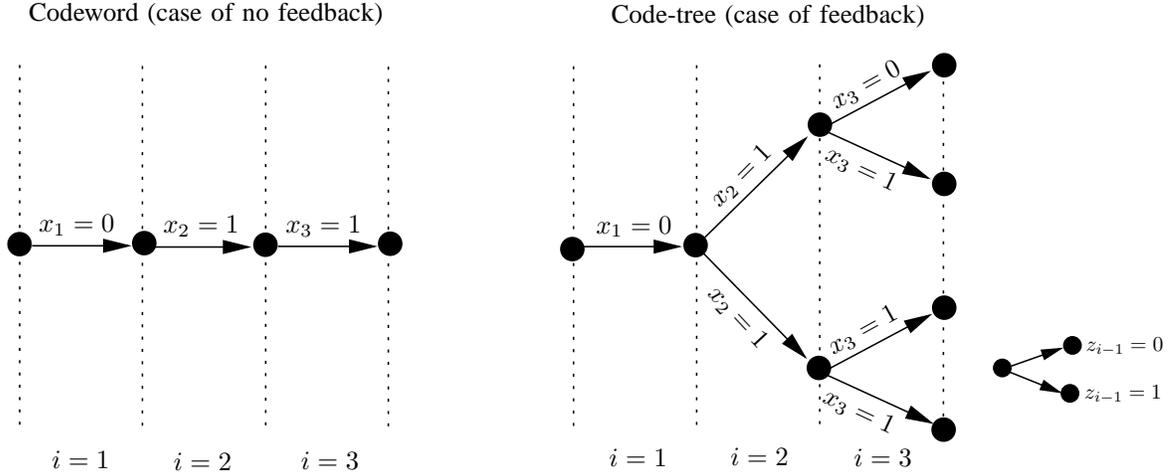}}
%\centerline{\includegraphics[width=16cm]{sd1.eps}}
\caption{Illustration of coding scheme for setting without
feedback and for setting with feedback. In the case of no feedback
each message is mapped to a codeword, and in the case of feedback
each message is mapped to a code-tree.} \label{f_codetree} }
\end{figure}

In the case of no feedback, a
coding block of length $N$ is a mapping of each message $m$ to a
codeword of length $N$ and is denoted by $x^N(m)$. In the case of
feedback, a coding block is a vector function whose $i^{th}$
component is a function of $m$ and the first $i-1$ components of
the received feedback. The mapping of the message $m$ and the
feedback $z^{i-1}$ to the input of the channel $x_i(m,z^{i-1})$ is
called a {\it code-tree} \cite[Ch. 9]{Blahut87} or {\it strategy}
\cite{Shannon61}. Figure \ref{f_codetree} shows an example of a
codeword of length $N=3$ for the case of no feedback and a
code-tree of depth $N=3$ for the case of binary feedback.

{\bf Randomly chosen coding scheme}:
 We choose the  $i^{th}$ channel input symbol
$x_i(m,z^{i-1})$ of the codeword $m$ by using a probability mass
function (PMF) based on previous symbols of the code
$x^{i-1}(m,z^{i-2})$ and previous feedback symbols $z^{i-1}$. The
first channel input symbol of codeword $m$ is chosen by the
probability function $Q(x_1)$. The second symbol of codeword $m$
is chosen for all possible feedback observations $z_1$ by the
probability function $Q(x_2|x^1,z^1)$ . The $i^{th}$ bit is chosen
for all possible $z^{i-1}$ by the probability function
$Q(x_i|x^{i-1},z^{i-1})$. This scheme of communication assumes
that the probability assignment of $x_i$ given $x^{i-1}$ and
$z^{i-1}$ cannot depend on $y^{i-1}$, because it is unavailable.
Therefore
\begin{equation}\label{e_as}
P(x_i|x^{i-1},z^{i-1}(y^{i-1}),y^{i-1})=P(x_i|x^{i-1},z^{i-1}(y^{i-1})),
\end{equation}

We also define $Q(x^N||z^{N-1})$, similarly as in
(\ref{e_causal_cond_delay_def}), to be the causal conditioning
probability
 \begin{equation}\label{e_def2}
Q(x^N||z^{N-1}) \triangleq \prod_{i=1}^{N} Q(x_i|x^{i-1},z^{i-1}).
\end{equation}

{\bf Encoding Scheme}: Each message $m$ has a code-tree.
Therefore, for any feedback $z^{N-1}$ and message $m$ there is a
unique input $x^N(m,z^{N-1})$ that was chosen randomly as
described in the previous paragraph. After choosing the coding
scheme, the decoder is made aware of the code-trees for all
possible messages. In our coding scheme the input $x^N(m,z^{N-1})$
is always a function of the message $m$ and the feedback, but in
order to make the equations shorter we also use the abbreviated
notation $x^N$ for $x^N(m,z^{N-1})$.

 {\bf Decoding Scheme} The decoder in our scheme is the Maximum likelihood (ML) decoder. Since the codewords depend on the feedback, two
different messages can have the same codeword for two different
outputs, therefore the regular ML $\arg \max_{x^N} P(y^N|x^N)$
cannot be used for decoding the message. Instead, the ML decoder
should be $\arg \max_{m} P(y^N|m)$ where $N$ is the block length.
The following equation shows that finding the most likely message
$m$ can be done by maximizing the causal conditioning
$P(y^N||x^{N})$:

\begin{equation}\label{e_ml0}
\arg \max_{m} \log P(y^N|m)=\arg \max_{m} \log P(y^N||x^{N}).
\end{equation}
The equality in (\ref{e_ml0}) is shown as follows:
\begin{eqnarray} \label{e_ml}
P(y^N|m) &= &  \prod_i P(y_i|y^{i-1},m) \nonumber \\
&\stackrel{(a)}{=}&  \prod_i P(y_i|y^{i-1},m,x^{i}(m,z^{i-1}(y^{i-1}))) \nonumber \\
&\stackrel{(b)}{=}&  \prod_i
P(y_i|y^{i-1},x^{i}(m,z^{i-1}(y^{i-1}))) \nonumber \\
&\stackrel{(c)}{=}& P(y^N||x^N).
\end{eqnarray}
Equality (a) holds because $x^{i}$ is uniquely determined by the
message $m$ and the feedback $z^{i-1}$, and the feedback $z^{i-1}$
is a deterministic function of $y^{i-1}$. Equality (b) holds
because according to the channel structure, $y_i$ does not depend
on $m$ given $x^i$. Equality (c) follows from the definition of
causal conditioning given in eq. (\ref{e_causal_cond_def}).

\subsection{ML decoding error bound} The next theorem, which is proved in Appendix \ref{a_t_MLB}, is a bound
on the expected ML decoding error probability with respect to the
random coding. Let $P_{e,m}$, as in \cite[Ch. 5.2]{Gallager68},
denote the probability of error using the ML decoder when message
$m$ is sent. When the source produces message $m$, there is a set
of outputs denoted by ${Y_m}^c$ that cause an error in decoding
the message $m$, i.e.,
\begin{equation}
P_{e,m}=\sum_{y^N \in {Y_m}^c}P(y^N|m).
\end{equation}
\begin{theorem} \label{t_MLB}
Suppose that an arbitrary message $m, 1\leq m \leq M$, enters the
encoder with feedback and that ML decoding is employed. Then the
average probability of decoding error over this ensemble of codes
is bounded, for any choice of $\rho,0 < \rho \leq 1$, by
\begin{equation}\label{e_MLB}
\mathbf E(P_{e,m}) \leq (M-1)^{\rho} \sum_{y^N} \left[ \sum_{x^N}
Q(x^N||z^{N-1}) P(y^N||x^N)^{\frac{1}{(1+\rho)}} \right]
^{1+\rho},
\end{equation}
where the expectation is with respect to the randomness in the
ensemble.
\end{theorem}

Let us define $P_{e,m}(s_0)$ to be the probability of error given
that the initial state of the channel is $s_0$ and message $m$ was
sent. The following theorem, which is proved in Appendix
\ref{a_t_fn}, establishes the existence of a code such that
$P_{e,m}(s_0)$ is small for all $1\leq m \leq M$.
\begin{theorem} \label{t_fn}
For an arbitrary finite-state channel with $|\mathcal{S}|$ states,
for any positive integer $N$ and any positive $R$, there exists an
$(N,M)$ code for which for all messages $m$, $1\leq m \leq
M=\lfloor 2^{NR}\rfloor$, all initial states $s_0$, and all
$\rho$, $0 \leq \rho \leq 1$, its probability of error is bounded
as
\begin{equation}\label{e_pfn}
P_{e,m}(s_0)\leq4|\mathcal{S}|2^{\{-N[-\rho R+F_N(\rho)]\}},
\end{equation}
where
\begin{equation} \label{eq_fn}
F_N(\rho)= \frac{\rho \log |\mathcal{S}|}{N}+\max_{
Q(x^N||z^{N-1})}\left[ \min_{s_0} E_{o,N}(\rho,
Q(x^N||z^{N-1}),s_0)\right],
\end{equation}
\begin{equation}\label{eq_E}
E_{o,N}(\rho,Q(x^N||z^{N-1}),s_0)=-\frac{1}{N}\log \sum_{y^N}
\left[ \sum_{x^N} Q(x^N||z^{N-1})
P(y^N||x^N,s_0)^{\frac{1}{(1+\rho)}} \right] ^{1+\rho}.
\end{equation}
\end{theorem}

The following theorem presents a few properties of the function
$E_{o,N}(\rho,Q(x^N||z^{N-1}),s_0)$ which is defined in eq.
(\ref{eq_E}), such as positivity of the function and its
derivative, and convexity of the function with respect to $\rho$.
\begin{theorem} \label{t_der}
The term $E_{o,N}(\rho,Q(x^N||z^{N-1}),s_0)$ has the following
properties:

\begin{equation}\label{eq_e1}
E_{o,N}(\rho,Q(x^N||z^{N-1}),s_0) \geq 0; \quad \rho \geq 0.
\end{equation}

\begin{equation}\label{eq_e2}
 \frac{1}{N}\mathcal{I}(Q(x^N||y^{N-1}),P(y^N||x^N,s_0)) \geq\frac{\partial
E_{o,N}(\rho, Q(x^N||z^{N-1}),s_0)}{\partial \rho} > 0; \quad \rho
\geq 0.
\end{equation}

\begin{equation}
\frac{\partial^2 E_{o,N}(\rho,Q(x^N||z^{N-1}),s_0)}{\partial
\rho^2} > 0; \quad \rho \geq 0.
\end{equation}
Furthermore, equality holds in (\ref{eq_e1}) when $\rho=0$, and
equality holds on the left side of eq. (\ref{eq_e2}) when
$\rho=0$.\end{theorem}
The proof of the theorem is omitted because
it is the same proof as Theorem 5.6.3 in \cite{Gallager68}.
Theorem 5.6.3 in \cite{Gallager68} states these same properties
with $Q(x^{N}||z^{N-1})$ and $P(y^N||x^N)$ replaced by $Q(x^{N})$
and $P(y^N|x^N)$, respectively. The proof of those properties only
requires that $\sum_{x^N} Q(x^{N}||z^{N-1})=1$ and $\sum_{x^N,y^N}
Q(x^{N}||z^{N-1}) P(y^N||x^N,s_0)=1$, which hold according to
Lemmas \ref{l_sumQ} and \ref{l_QP}. By using Lemma \ref{l_dir_mut}
we can substitute $\mathcal{I}(Q(x^N||y^{N-1}),P(y^N||x^N,s_0))$
in (\ref{eq_e2}) by the directed mutual information
$I(X^N\rightarrow Y^N|s_0)$.

\begin{lemma} \label{l_fn}
{\it Super additivity of $F_N(\rho)$.} For any given finite-state
channel, $F_N(\rho)$, as given by eq. (\ref{eq_fn}), satisfies
\begin{equation}
F_N(\rho)\geq \frac{n}{N} F_n(\rho) + \frac{l}{N}F_l(\rho)
\end{equation}
for all positive integers $n$ and $l$ with $N=n+l$.
\end{lemma}
The proof of the lemma is given in Appendix \ref{a_l_fn}.

\begin{lemma} \label{l_inf}{\it Convergence of $F_N(\rho)$.}
Let
\begin{equation}
F_\infty(\rho)=\sup_N F_N(\rho),
\end{equation}
then
\begin{equation}
\lim_{N \rightarrow \infty} F_N(\rho) =F_\infty(\rho),
\end{equation}
for $0 \leq \rho \leq 1$. Furthermore, the convergence is uniform
in $\rho$ and $F_\infty(\rho)$ is uniformly continuous for
$\rho\in[0,1]$.
\end{lemma}

\begin{proof}
Lemma 4A.2 in \cite{Gallager68} states that if a series $a_n$ is
super additive, i.e. $a_N\geq
\frac{n}{N}a_n+\frac{N-n}{N}a_{N-n}$, then $\lim_{N \rightarrow
\infty} a_N =\sup_N{a_N}$. Based on Lemma \ref{l_fn}, which states
that $\{F_N(\rho)\}$ is super additive, we get that $F_N(\rho)$
converges to $\sup_N F_N(\rho)$. From Theorem \ref{t_der} it
follows that
\begin{equation}\label{e_E0bound}
0\leq \frac{\partial E_{0,N}(\rho,Q(x^N||z^{N-1}),s_0)}{\partial
\rho} \leq \frac{1}{N}I(X^N\rightarrow Y^N)\leq \log
|\mathcal{Y}|,
\end{equation}
where $|\mathcal{Y}|$ is the size of the output alphabet. Using
this bound with the definition of $F_N$ given in eq. (\ref{eq_fn})
 we can bound the difference of $F_N(\rho)$ for any $0\leq \rho1 < \rho2 \leq
 1$ as
\begin{equation}\label{e_b1}
\frac{-(\rho_2-\rho_1)\log |\mathcal{S}|}{N}\leq
F_N(\rho_2)-F_N(\rho_1)\leq (\rho_2-\rho_1)\log |\mathcal{Y}|.
\end{equation}
A consequence of (\ref{e_b1}) is that the function $F_N(\rho)$ and
its slope are bounded independent of $N$ for each $0\leq \rho \leq
1$. Therefore the convergence is uniform in $\rho$ and $F_\infty$
is uniformly continuous.
\end{proof}

\begin{theorem}\label{t_c_conv}
Let us define
 \begin{equation}
 \underline C_N = \frac{1}{N}  \max_{Q(x^N||z^{N-1})} \min_{s_0}I(X^N
\rightarrow Y^N |s_0)
 \end{equation}
and
 \begin{equation}\label{e_underlineC}
 \underline{C} = \lim_{N \rightarrow \infty} \underline C_N.
 \end{equation}
Then, for a finite state channel with $|\mathcal{S}|$ states the
limit in \ref{e_underlineC} exists and
 \begin{equation}
  \lim_{N \rightarrow \infty} \underline C_N= \sup_{N}\left[ \underline C_N - \frac{log |\mathcal{S}|}{N}
  \right]=\sup_{N} \underline C_N.
 \end{equation}
\end{theorem}
\begin{proof}

Let us divide the input $x^N$ into two sets,  ${\bf x_1} = x_1^n$
and ${\bf x_2}= x_{n+1}^N$. Similarly, let us divide the output
$y^N$ into two sets ${\bf y_1} = y_1^n$ and ${\bf y_2}=
y_{n+1}^N$. Let $Q_n({\bf x_1}||{\bf z_1})=\prod_{i=1}^{n}
P(x_i|x^i,y^{i-1})$ and $Q_l({\bf x_2}||{\bf
z_2})=\prod_{i=1}^{l}P(x_{n+i}|x_{n+1}^{n+i},y_{n+1}^{n+i-1})$ be
the probability assignments that achieve $\underline C_n$ and
$\underline C_l$, respectively. Let us consider the probability
assignment $Q(x^N||z^{N-1})=Q_n({\bf x_1}||{\bf z_1})Q_l({\bf
x_2}||{\bf z_2})$. Then

\begin{eqnarray}
N \underline C_N & \geq & \min_{s_0} I(X^N \rightarrow Y^N|s_0) \nonumber \\
& \stackrel{(a)}{=} & \min_{s_0} \left[ \sum_{i=1}^{n} I(Y_i;X^i|Y^{i-1},s_0)+ \sum_{j=n+1}^{n+l}I(Y_j;X^j|Y^{j-1},s_0)   \right] \nonumber \\
%& \geq & \min_{s_0} \left[ \sum_{i=1}^{n} I(Y_i;X^i|Y^{i-1},s_0)+ \sum_{j=n+1}^{n+l}I(Y_j;X^j|Y^{j-1},s_0)   \right] \nonumber \\
& \stackrel{(b)}{\geq} & n\underline C_n +  \min_{s_0} \sum_{j=n+1}^{n+l}I(Y_j;X_{n+1}^j|Y_{n+1}^{j-1},{\bf y_1},s_0)  \nonumber \\
& \stackrel{(c)}{\geq} & n\underline C_n +  \min_{s_0} \sum_{j=n+1}^{n+l}I(Y_j;X_{n+1}^j|Y_{n+1}^{j-1},{\bf y_1},S_n,s_0)+\log |\mathcal{S}|  \nonumber \\
%& \stackrel{}{\geq} & n\underline C_n +  \min_{s_0} \sum_{j=n+1}^{n+l}I(Y_j;X_{n+1}^j|Y_{n+1}^{j-1},{\bf y_1},S_n)+\log |\mathcal{S}|  \nonumber \\
& \stackrel{}{\geq} & n\underline C_n +  \min_{s_0} \sum_{s_n} P(s_n|s_0) \sum_{j=n+1}^{n+l}I(Y_j;X_{n+1}^j|Y_{n+1}^{j-1},{\bf y_1},s_n)+\log |\mathcal{S}|  \nonumber \\
& \stackrel{}{\geq} & n\underline C_n +  \min_{s_n} \sum_{j=n+1}^{n+l}I(Y_j;X_{n+1}^j|Y_{n+1}^{j-1},s_n)+\log |\mathcal{S}|  \nonumber \\
& \stackrel{}{=} & n\underline C_n +  l\underline C_l+\log
|\mathcal{S}|.
\end{eqnarray}
Equality (a) is due to the definition of the directed information.
Inequality (b) holds because $\underline C_n$ is the first term
and for the second term we use the fact that $I(X;Y,Z)\geq I(X;Y)$
for any random variables $(X,Y,Z)$. Inequality (c) is due to Lemma
\ref{l_dirB}. Rearranging  the inequality we get:
\begin{eqnarray}
N \left[\underline C_N -\frac{\log |\mathcal{S}|}{N}\right] \geq
n\left[\underline C_n - \frac{\log |\mathcal{S}|}{n}\right] +
l\left[\underline C_l-\frac{\log |\mathcal{S}|}{l}\right].
\end{eqnarray}
Finally, by using the convergence of a super additive sequence,
the theorem is proved.
\end{proof}

A rate $R$ is said to be {\it achievable} if there exists a
sequence of block codes $(N,\lceil 2^{NR} \rceil)$ such that the
maximal probability of error $\max_m P_{e,m}(s_0)$ tends to zero
as $N \to \infty$ for all initial states $s_0$ \cite{CovThom}. The
following theorem states that any rate $R$ that satisfies
$R<\underline C$ is achievable.

\begin{theorem}\label{t_ach}
For any given finite-state channel , let
\begin{equation}
E_r(R)=\max_{0\leq \rho \leq 1}[F_\infty(\rho)-\rho R].
\end{equation}
Then, for any $\epsilon>0$, there exists $N(\epsilon)$ such that
for $N\geq N(\epsilon)$  there exists an $(N,M)$
 code such that for all $m,1\leq m \leq M=\lceil 2^{NR}\rceil$, and all initial
 states,
 \begin{equation}\label{e_pemE}
 P_{e,m}(s_0)\leq 2^{-N[E_r(R)-\epsilon]}.
 \end{equation}
Furthermore, for $0\leq R < \underline C $, $E_r(R)$ is strictly
positive, and therefore the error can be arbitrarily small for $N$
large enough.
\end{theorem}

\begin{proof}
For any rate $R$, we can rewrite eq. (\ref{e_pfn}) as
\begin{equation}\label{e_peFn}
P_{e,m}(s_0)\leq 2^{-N(-\rho R +F_N(\rho)-\frac{\log
4|\mathcal{S}|}{N})}.
\end{equation}
Because of the uniform convergence in $\rho$ proven in Lemma
\ref{l_inf}, for all $\epsilon>0$, there exists an $N(\epsilon)$
that does not depend on $\rho$ such that, for $N\geq N(\epsilon)$,
\begin{equation}
F_{\infty}(\rho)-F_N(\rho)+\frac{\log 4|\mathcal{S}|}{N}\leq
\epsilon; \qquad 0\leq \rho \leq 1.
\end{equation}

Hence, it follows from (\ref{e_peFn}) that
\begin{equation}\label{e_peFinfty}
P_{e,m}(s_0)\leq 2^{-N(-\rho R +F_{\infty}(\rho)-\epsilon)}.
\end{equation}
If we choose the $\rho$ that maximizes $-\rho R +F_{\infty}(\rho)$
(note that $F_\infty(\rho)$ and therefore $-\rho
R+F_{\infty}(\rho)$ is continuous in $\rho\in[0,1]$, so there
exists a maximizing $\rho$), then inequality (\ref{e_peFinfty})
becomes inequality (\ref{e_pemE}), proving the first part of the
theorem.

Now let us show that if $R< \underline C$, then $E_r(R)>0$, which
will prove the second part of the theorem. Let us define
$\delta\triangleq \underline C - R$ . According to Theorem
\ref{t_c_conv}, $C_N$ converges to $\sup_N C_N=\underline C$,
hence we can choose $N$ large enough so that the following
inequality holds:
\begin{equation}
 \underline C_N \geq R+\frac{\log |\mathcal{S}|}{N}+
 \frac{\delta}{2}.
\end{equation}
From Theorem \ref{t_der},  we have
\begin{equation}
\frac{\partial E_{o,N}(\rho,Q (x^N||z^{N-1}),s_0)}{\partial \rho}
\leq \underline C_N, \qquad \forall s_0,
\end{equation}
where $Q (x^N||z^{N-1})$ is chosen to be the distribution that
achieves  $\underline C_N$.

Note that $E_{o,N}(\rho,Q (x^N||z^{N-1}),s_0)$ is zero when
$\rho=0$, is a continuous function of $\rho$, and the derivative
at zero with respect to $\rho$ is equal to $\underline C_N \geq
R+\frac{\log |\mathcal{S}|}{N}+\frac{\delta}{2}$. Thus, for each
state $s_0$ there is a range $\rho>0$ such that
\begin{equation}\label{e_EoN}
E_{o,N}(\rho,Q (x^N||z^{N-1}),s_0) - \rho(R+\frac{\log
|\mathcal{S}|}{N}) > 0.
\end{equation}
Moreover, because the number of states is finite, there exists a
$\rho^*>0$ for which the inequality (\ref{e_EoN}) is true for all
$s_0$. Thus,
\begin{equation}
F_\infty(\rho^*) \geq F_N(\rho^*) \geq E_{o,N}(\rho,Q
(x^N||z^{N-1}),s_0) - \rho^*\frac{\log |\mathcal{S}|}{N}>\rho^*R,
\qquad \forall s_0,
\end{equation}
and thus $E_r(R)>0$ for $R<\underline C$.
\end{proof}

\subsection{Feedback that is a deterministic function of a finite
tuple of the output\label{s_tuple}} We proved Theorem \ref{t_ach}
for the case when the feedback $z_i$ is a deterministic function
of the output at time $i$, i.e $z_i=z(y_i)$. We now extend the
theorem to the case where the feedback is a deterministic function
of a finite tuple of the output, i.e.
$z_i=z(y_{i-D-1},...,y_{i})$.

Consider the case $D=2$.
%Let $F$ that is FSC and with
%the feedback $z_i(y_{i-1},y_{i})$ assuming that at time $i=1$ the
%feedback is $z_i(y_{0},y_{i})$ when $y_{0}$ is chosen with some
%probability $P(y_0)$ independently of the initial state of the
%channel $s_0$, and it can be considered as the initial state of
%the feedback.
Let us construct a new finite state channel, with input $x_i$, and
output $\tilde y_i$ that is the tuple $\{y_{i-1},y_{i}\}$. The
state of the new channel $\tilde s_i$ is the tuple $\{s_i,y_i\}$.

Let us verify that the definition of a FSC holds for the new
channel:
\begin{eqnarray}
P(\tilde y_i,\tilde s_i|\tilde y^{i-1},\tilde s^{i-1},x^i)&=&P(y_i,y_{i-1},s_i,y_{i}|y^{i-1},s^{i-1},y^{i-1},x^i) \nonumber\\
&=&P(y_i,y_{i-1},s_i|y_{i-1},s_{i-1},x_i) \nonumber \\
 &=&P(\tilde y_i,\tilde s_i|\tilde  s_{i-1},x_i)
\end{eqnarray}
%The capacity of the new channel $F'$ and the original channel are
%the same because for every input $x^i$ we can construct $y^i$ from
%$y'^i$ and vice versa. The only difference is that the series
%$y'^i$ includes an additional element $y_0$ but because it has a
%finite cardinality it does not add to the information per bit when
%the sequence length goes to infinity.
Both channels are equivalent, and because the feedback $z_i$ is a
deterministic function of the output of the new channel, $\tilde
y_i$, we can apply Theorem \ref{t_ach} and get that any achievable
rate satisfies

\begin{eqnarray}
R &\leq & \underline C_N = \frac{1}{N}  \max_{Q(x^N||z^{N-1})}
\min_{\tilde s_0}I(X^N \rightarrow \tilde Y^N |\tilde s_0) \nonumber \\
&= &  \frac{1}{N}  \max_{Q(x^N||z^{N-1})} \min_{s_0,y_0}I(X^N
\rightarrow
\{Y^N,Y_0^{N-1} \}|s_0,y_0) \nonumber \\
&= &  \frac{1}{N} \max_{Q(x^N||z^{N-1})} \min_{s_0,y_0}
\sum_{i=1}^{N}I(X^i;Y_i,Y_{i-1}|Y^{i-1},Y^{i-2},s_0,y_0) \nonumber \\
&= &  \frac{1}{N} \max_{Q(x^N||z^{N-1})} \min_{s_0,y_0}
\sum_{i=1}^{N} H(Y_i,Y_{i-1}|Y^{i-1},Y^{i-2},s_0,y_0)- H(Y_i,Y_{i-1}|Y^{i-1},Y^{i-2},X^i,s_0,y_0)\nonumber \\
&= &  \frac{1}{N} \max_{Q(x^N||z^{N-1})} \min_{s_0,y_0}
\sum_{i=1}^{N} H(Y_i|Y^{i-1},s_0,y_0)- H(Y_i|Y^{i-1},X^i,s_0,y_0)\nonumber \\
&= &  \frac{1}{N} \max_{Q(x^N||z^{N-1})} \min_{s_0,y_0} I(X^N
\rightarrow
Y^N|s_0,y_0)\nonumber \\
\end{eqnarray}

This result can be extended by induction to the general case where
the feedback $z_i$ depends on a tuple of $D$ outputs, % which means
%that the feedback is a deterministic function
%$Z_i=k(y_{i-M+1}^{i})$ and the initial state of the feedback is
%$y_{2-M}, \cdots ,y_0$. We have proved that for the case where
%$M=2$ the channel is a FSC, hence we can now extend it to $M+1$ by
%the same way as we extended from $M=1$ to $M=2$. Finally, we get
%that the achievable rate is
leading to the achievability of any rate smaller than $\lim_{N \to
\infty} \underline C_N$, where in this setting
\begin{eqnarray}\label{e_ctuple}
\underline C_N =\frac{1}{N} \max_{Q(x^N||z^{N-1})}
 \min_{\{s_0,y_{2-M}, \cdots, y_0\}} I(X^N \rightarrow
Y^N|s_0,y_{2-M}, \cdots ,y_0).
\end{eqnarray}

\section{Upper bound on the feedback capacity\label{s_upper_bound_on_capacity}}
\begin{theorem} \label{t_Cupper}
The capacity of a channel where the input is $x^N$ and the output
is $y^N$ and the channel has a time invariant deterministic
feedback, as presented in Fig. \ref{f_1}, is upper bounded as
\begin{equation}\label{e_Cupper_bound}
C_{FB} \leq  \lim_{N \rightarrow \infty} \max _{Q(x^N||z^{N-1})}
I(X^N \rightarrow Y^N).
\end{equation}
\end{theorem}

\begin{proof}
Let $W$ be the message, chosen according to a uniform distribution
${\Pr(W=w)=2^{-NR}}$. The input to the channel $x_i$ is a function
of the message $W$ and the arbitrary deterministic feedback output
$z^{i-1}(y^{i-1})$. We have
\begin{eqnarray}\label{e_con}
NR &= & H(W) \nonumber \\
&=&I(W;Y^N)+H(W|Y^N) \nonumber \\
& \stackrel{(a)}{\leq} & I(Y^N;W) + 1 + P_2^{(N)}NR \nonumber \\
&=& H(Y^N)-H(Y^N|W)+1 + P_2^{(N)}NR \nonumber \\
&\stackrel{(b)}{=}& \sum_{i=1}^{N} H(Y_i|Y^{i-1}) - \sum_{i=1}^{N} H(Y_i|W,Y^{i-1})+1 + P_2^{(N)}NR \nonumber \\
&\stackrel{(c)}{=}& \sum_{i=1}^{N} H(Y_i|Y^{i-1}) - \sum_{i=1}^{N} H(Y_i|W,Y^{i-1},X^i(W,z^{i-1}(Y^{i-1})))+1+ P_2^{(N)}NR \nonumber \\
&\stackrel{(d)}=& \sum_{i=1}^{N} H(Y_i|Y^{i-1}) - \sum_{i=1}^{N} H(Y_i|Y^{i-1},X^i)+1+ P_2^{(N)}NR \nonumber \\
&=& \sum_{i=1}^{N} I(Y_i;X^i|Y^{i-1})+1+ P_2^{(N)}NR
\end{eqnarray}
Inequality (a) holds because of Fano's inequality. Equality (b)
holds because of the chain rule. Equality (c) holds because $x_i$
is a deterministic function given the message $W$ and the feedback
$z^{i-1}$, where the feedback $z^{i-1}$ is a deterministic
function of the output. Equality (d) holds because the random
variables $W,X_i,Y^{i-1},Y_i$ form the Markov chain
$W-(X_i,Y^{i-1})-Y_i$. By dividing both sides of the equation by
$N$, maximizing over all possible input distributions, and letting
$N \rightarrow \infty$ we get that in order to have an error
probability arbitrarily small, the rate $R$ must satisfy:
\begin{equation}
R \leq \lim_{N \rightarrow \infty} \frac{1}{N}
\max_{Q(X^N||Z^{N-1})} \sum_{i=1}^{N} I(Y_i;X^i|Y^{i-1})=\lim_{N
\rightarrow \infty} \max_{Q(X^N||Z^{N-1})} \frac{1}{N}
\sum_{i=1}^{N} I(X^N \to Y^N).
\end{equation}
This completes the proof.
\end{proof}

Remark: The converse proof is with respect to the average error
over all messages. This, of course, implies that it is also true
with respect to the maximum error over all messages. In the
achievability part we proved that the maximum error over all
messages goes to zero when $R\leq \underline C$ which, of course,
also implies that the average error goes to zero. Hence, both the
achievability and the converse are true with respect to average
error probability and maximum error probability over all messages.

\section{Indecomposable FSC without ISI \label{s_indeco}}
In this section we assume that the channel states evolve according
to a Markov chain which does not depend on the input, namely
$P(y_i,s_i|s_{i-1},x_i)=P(s_i|s_{i-1})P(y_i|s_i,s_{i-1},x_i)$. In
addition, we assume that the Markov chain is indecomposable. Such
a channel is called a Finite State Markovian indecomposable
channel (FSMIC) in \cite{Jelinek65}, however another suitable name
which we adopt henceforth is a FSC without ISI. The difference
between this channel and the indecomposable FSC defined in
\cite{Gallager68, Blackwell58} is that here we make an additional
assumption that the transition probability between states is not a
function of the input.

A Markov chain with transition matrix $P(i,j)$ is indecomposable
if it contains only one ergodic class \cite{Doob53}. An equivalent
definition is that the effect of the initial state of the Markov
chain dies away with time. More precisely:
\begin{definition} \label{d_ind}
A Markov Chain is indecomposable if, for every $\epsilon>0$, there
exists an $N_0$ such that for $N\geq N_0$,
\begin{equation}\label{e_ind_property}
|P(s_N|s_0)-P(s_N|s'_0)|\leq \epsilon
\end{equation}
for all $s_N,s_0,s'_0$.
\end{definition}

 In this section we prove that for a FSC without ISI the
achievable rate does not depend on the initial state $s_0$ and
therefore the lower bound and the upper bound on the capacity as
given in (\ref{e_underlineC}) and (\ref{e_Cupper_bound}) are
equal.

Let us define
 \begin{equation}
 \overline C_N = \frac{1}{N}  \max_{Q(x^N||z^{N-1})} \max_{s_0}I(X^N
\rightarrow Y^N |s_0)
 \end{equation}
and
 \begin{equation}\label{e_ovelineC_def}
 \overline{C} = \lim_{N \rightarrow \infty} \overline C_N,
 \end{equation}
 a limit that will be shown to exist.
In addition let us define
 \begin{equation}
 C = \lim_{N \rightarrow \infty} \frac{1}{N}  \max_{Q(x^N||z^{N-1})} I(X^N
\rightarrow Y^N ).
 \end{equation}

\begin{theorem}\label{t_Cunder_over_equal_no_isi}
For a FSC without ISI,
\begin{equation}
 \overline{C}=\underline{C}=C=\lim_{N \rightarrow \infty} \frac{1}{N}  \max_{Q(x^N||z^{N-1})} I(X^N
\rightarrow Y^N ),
 \end{equation}
 where $\overline{C}$ was defined in (\ref{e_ovelineC_def}) and
 $\underline{C}$ in (\ref{e_underlineC}).
\end{theorem}
\begin{proof}
For arbitrary $N$, let $Q_N(x^N||z^{N-1})$ and $s_0'$ be the input
distribution and the initial state that maximize $I_Q(X^N
\rightarrow Y^N|s_0')$ and let $s_0''$ denote the initial state
that minimizes $I_Q(X^N \rightarrow Y^N|s_0'')$ for the same input
distribution, where the subscript in $I_Q$ is added to emphasize
its dependence on $Q_N$, though we suppress the subscript $N$ from
$Q_N$. Thus, we have
\begin{equation}\label{e_Csandwitch}
  \frac{1}{N} I_Q(X^N \rightarrow
 Y^N|s_0') \stackrel{}{=}  \overline C_N \stackrel{}{\geq} \underline C_N  \stackrel{}  {\geq}  \frac{1}{N} I_Q(X^N \rightarrow
 Y^N|s_0'').
\end{equation}
The equation holds due to the definitions of $\overline C_N$ and
$\underline C_N$. Next, we will prove that $\lim_{N \rightarrow
\infty}\frac{1}{N} I_Q(X^N \rightarrow
 Y^N|s_0')=\lim_{N \rightarrow \infty}\frac{1}{N} I_Q(X^N \rightarrow
 Y^N|s_0'')$ and therefore  $\overline{C}=\underline{C}$.
%Equations \ref{e_in1}-\ref{e_in4} are true for any input
%distribution  Throught the remindar of the proof, the distribution
%of $(X^N,Y^N)$ should be understood as that is induced by $Q_N$.

 Let $n+l=N$, where $n$ and $l$ are positive integers and let
$s_0'$ and $s_0''$ be any two initial states. Let the random
variable $S_n$ be the state at time $n$. We would like to
emphasize that the difference in the letter case notation is
because $s_0'$ and $s_0''$ are specific states while $S_n$ is a
random variable. Then
\begin{eqnarray}\label{e_in1}
\lefteqn{\frac{1}{N} I_Q(X^N \rightarrow Y^N |s_0')
} \nonumber \\
&\stackrel{(a)}{\leq}& \frac{1}{N} [ \log |\mathcal{S}| + I_Q(X^N \rightarrow Y^N |s_0',S_n)] \nonumber \\
&=& \frac{1}{N} \left [ \log |\mathcal{S}| +
\sum_{i=1}^{n}I_Q(Y_i;X^i|Y^{i-1},S_n,s_0') +
\sum_{i=n+1}^{N}I_Q(Y_i;X_1^n,X_{n+1}^i|Y^{i-1},S_n,s_0')\right] \nonumber \\
& \stackrel{(b)}{\leq}& \frac{1}{N} \left[ \log |\mathcal{S}| +
n\log |\mathcal{Y}| +
\sum_{i=n+1}^{N}I_Q(Y_i;X_1^n,X_{n+1}^i|Y^{i-1},S_n,s_0')\right] \nonumber \\
&=& \frac{1}{N} \left[ \log |\mathcal{S}| + n\log |\mathcal{Y}| +
\sum_{i=n+1}^{N}[I_Q(Y_i;X_{n+1}^i|Y^{i-1},S_n,s_0')+I_Q(Y_i;X_1^n|Y^{i-1},X_{n+1}^i,S_n,s_0')] \right] \nonumber \\
& \stackrel{(c)}{=}& \frac{1}{N} \left[ \log |\mathcal{S}| + n\log
|\mathcal{Y}| + \sum_{i=n+1}^{N}
I_Q(Y_i;X_{n+1}^i|Y^{i-1},S_n,s_0') \right].
\end{eqnarray}
Inequality (a) is due to Lemma \ref{l_dirB}. Inequality (b) is due
to the bound $I_Q(Y_i;X^i|Y^{i-1},S_n,s_0')\leq \log
|\mathcal{Y}|$. Equality (c) holds because given the state $S_n$
and the input after time $n$, the output after time $n$ does not
depend on the input before time $n$, i.e.
$P(y_i|y^{i-1},x_{n+1}^i,s_n,x_1^n,s_0)=P(y_i|y^{i-1},x_{n+1}^i,s_n,s_0)$
, $i>n$. By using inequality (\ref{e_in1}) we can bound the
difference between the directed information starting at two
different states:

\begin{eqnarray}\label{e_in2}
\lefteqn{\frac{1}{N}| I_Q(X^N \rightarrow Y^N |s_0') - I_Q(X^N
\rightarrow Y^N
|s_0'')|}\nonumber \\
& \leq & \frac{1}{N} \left [ \log |\mathcal{S}| + n\log
|\mathcal{Y}| +\sum_{i=n+1}^{N}[
I_Q(Y_i;X_{n+1}^i|Y^{i-1},S_n,s_0')-
I_Q(Y_i;X_{n+1}^i|Y^{i-1},S_n,s_0'')] \right]
\end{eqnarray}
The sum in the last inequality can be bounded by using the
indecomposability property of the Markov chain. For every $i>n$ we
have:
\begin{eqnarray}\label{e_in3}
\lefteqn{I_Q(Y_i;X_{n+1}^i|Y^{i-1},S_n,s_0')-
I_Q(Y_i;X_{n+1}^i|Y^{i-1},S_n,s_0'')} \nonumber \\
&\stackrel{(a)}{=}&
\sum_{s_n}[P(s_n|s_0')-P(s_n|s_0'')]I_Q(Y_i;X_{n+1}^i|Y_{i-1}^{i-1},s_n,s_0')
\nonumber \\
&\stackrel{(b)}{\leq}&  \sum_{s_n}[P(s_n|s_0')-P(s_n|s_0'')] \log |\mathcal{Y}|\nonumber \\
&\stackrel{(c)}{\leq}&  \epsilon_n \log |\mathcal{Y}|.
%&\stackrel{}{=}&  \sum_{s_n} \sum_{x_n}[P(s_n,x_n|s_0')-P(s_n,x_n|s_0'')] \log B\nonumber \\
%&\stackrel{}{=}&  \sum_{s_n} \sum_{x_n}[P(x_n|s_0')P(s_n|s_0',x_n)-P(x_n|s_0'')P(s_n|s_0'',x_n))] \log B\nonumber \\
%&\stackrel{}{\leq}&  \sum_{s_n} \sum_{x_n}[P(x_n|s_0')(s_n|s_0',x_n)-P(s_n|s_0'',x_n)] \log B\nonumber \\
\end{eqnarray}
Equality (a) is achieved by summing over all possible states
$s_n$. Inequality (b) is achieved by bounding the magnitude of
each term in the sum by $\log |\mathcal{Y}|$. Inequality (c) holds
by defining $\epsilon_n\triangleq
\max_{s_0',s_0''}|\sum_{s_n}P(s_n|s_0')-P(s_n|s_0'')|$. Combining
eq. (\ref{e_in2}) and eq. (\ref{e_in3}) we obtain:
\begin{eqnarray}\label{e_in4}
\frac{1}{N}| I_Q(X^N \rightarrow Y^N |s_0') -  I_Q(X^N \rightarrow
Y^N |s_0'')|& \leq & \frac{1}{N} [ \log |\mathcal{S}| + n\log
|\mathcal{Y}| +\epsilon_n\cdot |\mathcal{S}| \cdot l \log
|\mathcal{Y}|]\nonumber \\
& \leq & \frac{1}{N}  \log |\mathcal{S}| + \frac{n}{N}\log
|\mathcal{Y}| +\epsilon_n \log |\mathcal{Y}|]\nonumber .
\end{eqnarray}
Since, by the indecomposability of the channel, $\epsilon\to
\infty$, and since inequality (\ref{e_in4}) holds for all $0\leq
n\leq N$, it follows from inequality (\ref{e_in4}) (by letting $n$
increase without bound, but sub-linearly in $N$) that
\begin{eqnarray}\label{e_in5}
\lim_{N \rightarrow \infty} \frac{1}{N}| I_Q(X^N \rightarrow Y^N
|s_0') - I_Q(X^N \rightarrow Y^N |s_0'')|& = & 0.
\end{eqnarray}

Up to now, we have proved that $\lim_{N\to \infty} \underline
C_N=\lim_{N\to \infty} \overline C_N$ and this is because of eq.
(\ref{e_in5}) and (\ref{e_Csandwitch}). Finally, we show that even
without conditioning on $s_0$ we get the same limit. Indeed,
\begin{eqnarray}\label{e_in6}
C_N \triangleq  \frac{1}{N}\max_{Q(x^N||z^{N-1})}I(X^N \rightarrow
Y^N ) &\stackrel{(a)}{=}& \frac{1}{N}\max_{Q(x^N||z^{N-1})}I(X^N
\rightarrow Y^N |S_0) +\frac{|\mathcal{S}|}{N}\nonumber \\
&=&  \frac{1}{N}\max_{Q(x^N||z^{N-1})} \sum_{s_0}P(s_0) I(X^N
\rightarrow Y^N |s_0)  +\frac{|\mathcal{S}|}{N}\nonumber\\
%&=& \sum_{s_0}P(s_0)  \frac{1}{N}\max_{Q(x^N||z^{N-1})} I(X^N
%\rightarrow Y^N |s_0)  +\frac{|\mathcal{S}|}{N}\nonumber\\
&\stackrel{}{\leq}&  \frac{1}{N}\max_{Q(x^N||z^{N-1})} \max_{s_0}
I(X^N \rightarrow Y^N |s_0) +\frac{|\mathcal{S}|}{N}\nonumber \\
&\stackrel{}{=}&  \overline C_N +\frac{|\mathcal{S}|}{N}.
\end{eqnarray}
Equality (a) holds because, according to Lemma \ref{l_dirB}, the
magnitude of the difference between the expression in the two
sides of the equation is bounded by $\frac{|\mathcal{S}|}{N}$. In
a similar way we prove that $C_N\geq \underline
C_N-\frac{|\mathcal{S}|}{N}$ and therefore we get that $\lim_{N\to
\infty} \underline C_N= \lim_{N\to \infty} C_N=\lim_{N\to \infty}
\overline C_N$ which concludes the proof.
\end{proof}

The {\it capacity} of a channel is defined as the supremum over
all achievable rates, analogous to what is done in the absence of
feedback\cite{CovThom}.
\begin{theorem} \label{t_capacity_FSC_no_ISI}
The capacity of an Indecomposable FSC without ISI with a time
invariant feedback $z(y_i)$ is given by
 \begin{equation}\label{e_t_capacity_FSC_no_ISI}
 C = \lim_{N \rightarrow \infty} \frac{1}{N}  \max_{Q(x^N||z^{N-1})} I(X^N
\rightarrow Y^N ),
 \end{equation}
where $C$ denotes the capacity of the channel in the presence of
feedback.
\end{theorem}

\begin{proof}
According to Theorem \ref{t_ach}, for any given finite state
channel, any rate $R$ in the range $0\leq R< \underline C$, is
achievable. According to Theorem \ref{t_Cupper}, the upper bound
on capacity of a FSC is $\lim_{N \rightarrow \infty} \frac{1}{N}
\max_{Q(x^N||z^{N-1})} I(X^N \rightarrow Y^N )$. Hence we get that
the capacity $C$ is bounded from below and from above by:
\begin{equation}
\underline C  \leq  C\leq \lim_{N \rightarrow \infty} \frac{1}{N}
\max_{Q(x^N||z^{N-1})} I(X^N \rightarrow Y^N ).
\end{equation}
Theorem \ref{t_Cunder_over_equal_no_isi} states that, for an
indecomposable FSC without ISI, the upper bound equals the lower
bound, i.e. $\underline C = \lim_{N \rightarrow \infty}
\frac{1}{N} \max_{Q(x^N||z^{N-1})} I(X^N \rightarrow Y^N ) $, and
therefore the capacity is given by
(\ref{e_t_capacity_FSC_no_ISI}).
 \end{proof}

\begin{figure}{
 \psfrag{v1\r}[][][0.8]{$m$}\psfrag{w1\r}[][][0.8]{Message}
 \psfrag{u1\r}[][][0.8]{Encoder}
\psfrag{d1\r}[][][0.8]{$x_i(m,z^{i-1})$}
\psfrag{v2\r}[][][0.8]{$x_i$} \psfrag{w2\r}[][][0.8]{$$}
 \psfrag{u2\r}[][][0.8]{Finite State Channel}
\psfrag{d2\r}[][][0.8]{$P(l_i,y_i,s_i|x_i,s_{i-1})$}
\psfrag{v3\r}[][][0.8]{$y_i,l_i \ \ $}
\psfrag{w3\r}[][][0.8]{$$}\psfrag{v1\r}[][][0.8]{$m$}
\psfrag{u3\r}[][][0.8]{Decoder} \psfrag{d3\r}[][][0.8]{$\hat
m(y^N,l^N)$} \psfrag{u5\r}[][][0.8]{Feedback Generator}
\psfrag{v5\r}[][][0.8]{$z_i(y_{i},l_i)\ $}
\psfrag{v4\r}[][][0.8]{$y_i,l_i\ \ $} \psfrag{w4\r}[][][0.8]{$$}
\psfrag{u4\r}[][][0.8]{Unit Delay}
\psfrag{D5\r}[][][0.8]{Time-Invariant}
\psfrag{D6\r}[][][0.8]{Function}
\psfrag{v7\r}[][][0.8]{$z_{i-1}(y_{i-1},l_{i-1})\ \ \ \ $}
 \psfrag{w5\r}[][][0.8]{$$}
\psfrag{v6\r}[][][0.8]{$\hat m$}\psfrag{w6a\r}[][][0.8]{Estimated}
\psfrag{w6b\r}[][][0.8]{Message} \centering
\includegraphics[width=16cm]{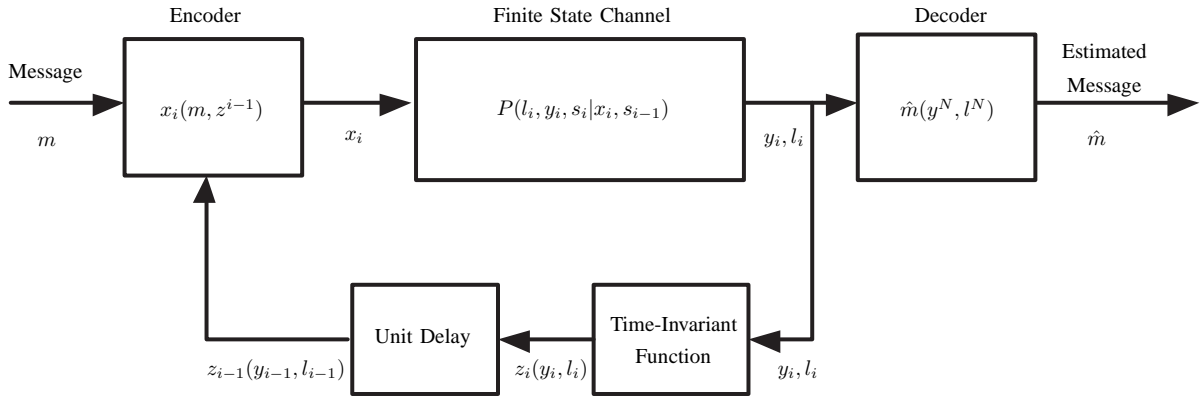}
\centering \caption{Channel with feedback and side information
$l_i$. } \label{f_3}
}\end{figure} %%when I save on visio I used AI 3 - and I had to restart the computer and not all visio features work with AI3

\section{Feedback and side information \label{s_feedbacz_and_side_information}}
The results of the previous sections can be extended to the case
where side information is available at the decoder that might be
also fed back to the encoder. Let $l_i$ be the side information
available at the decoder and the setting of communication the one
in Fig. \ref{f_3}. If the side information $l_i$ satisfies
\begin{equation}\label{e_li}
P(l_i,y_i,s_i|s^{i-1},x^{i},y^{i-1},l^{i-1})=P(l_i,y_i,s_i|s_{i-1},x_{i})
\end{equation}
then it follows that
\begin{equation}\label{e_ybar}
P(\bar y_i,s_i | s^{i-1},x^{i},{\bar y}^{i-1})=P(\bar
y_i,s_i|s_{i-1},x_{i}),
\end{equation}
where $\bar y_i=(l_i,y_i)$. We can now apply Theorem \ref{t_ach}
and get:
\begin{eqnarray}\label{e_cli}
  \underline C_N &= &\frac{1}{N}  \max_{Q(x^N||z^{N-1})} \min_{s_0}I(X^N
\rightarrow \{Y^N,L^N\} |s_0),
% = \frac{1}{N}  \max_{Q(x^N||z^{N-1})}
%\min_{s_0}I(X^N \rightarrow \bar Y^N |s_0) \nonumber \\
%&= &  \frac{1}{N} \max_{Q(x^N||z^{N-1})} \min_{s_0}
%\sum_{i=1}^{N}I(X^i;\bar Y_{i}|\bar Y^{i-1},s_0) \nonumber \\
%&= &  \frac{1}{N} \max_{Q(x^N||z^{N-1})} \min_{s_0}
%\sum_{i=1}^{N}I(X^i; Y_{i},L_{i}|Y^{i-1}, L^{i-1},s_0)
%\nonumber \\
\end{eqnarray}
where $z_{i-1}$ denotes the feedback available at the receiver at
time $i$ which is a time-invariant function of $l_{i-1}$ and
$y_{i-1}$.

While many cases of side information can be studied, we are going
to consider only the case in which the side information is the
state of the channel, i.e. $l_i=s_i$, which is fed back to the
encoder, namely we let $z_i(y_i,l_i)=s_i$. In this section we no
longer assume that there is no ISI, instead we assume that the FSC
is {\it strongly connected}, which we defines as follows.
\begin{definition}
We say that a finite state channel is strongly connected if there
exists an input distribution $\{Q(x_t|s_{t-1})\}_{t\geq 1}$ and
integer $T$ such that %the probability that the channel reaches
%state $s$ from any starting state $s'$, in less then $T$
%time-steps, is positive.
\begin{equation}
\Pr \{S_t=s \text{ for some } 1\leq t \leq T|S_0=s'\}>0, \ \forall
s',s.
\end{equation}
\end{definition}

\begin{theorem}\label{t_feedbacz_does_not}
Feedback does not increase the capacity of a strongly connected
FSC when the state of the channel is known both at the encoder and
the decoder. Furthermore, the capacity of the channel under this
setting is given by
\begin{equation}\label{e_css}
 C \stackrel{}{=}\lim_{n\to \infty} \frac{1}{N}
\max_{\{Q(x_i|s_{i-1})\}} \sum_{i=1}^{N} I(
Y_{i},S_i;X_i|S_{i-1}).
\end{equation}
\end{theorem}
A straightforward consequence of this theorem is that feedback
does not increase the capacity of a discrete memoryless channel
(DMC), which Shannon proved in 1956 \cite{shannon56}. A DMC can be
considered as an FSC with only one state, and therefore the state
of the channel is known to the encoder and the decoder.

\begin{proof}
First, we notice that because the state of the channel is known
both to the encoder and the decoder, and because the FSC is
strongly connected, we can assume that with probability
$1-\epsilon$, where $\epsilon$ is arbitrarily small, the FSC
channel can be driven, in a finite time, to the state that
maximizes the achievable rate. Hence, the achievable rate does not
depend on the initial state and the capacity of the channel in the
present of feedback, which we denote as $C^{(F)}$, is given by
$\lim_{N\to \infty} C_N^{(F)}$, where $C_N^{(F)}$ satisfies
\begin{eqnarray}\label{e_cli4}
C_N^{(F)} &\stackrel{}{=}& \frac{1}{N} \max_{Q(x^N||z^{N-1})}
I(X^N \rightarrow \{Y^N,L^N\})
 \nonumber \\
&\stackrel{(a)}{=} &\frac{1}{N}
\max_{Q(x^N||z^{N-1})} \sum_{i=1}^{N}I(X^i; Y_{i},S_{i}|Y^{i-1}, S^{i-1}) \nonumber \\
&\stackrel{}{=} &\frac{1}{N}
\max_{Q(x^N||z^{N-1})} \sum_{i=1}^{N}H( Y_{i},S_i|Y^{i-1}, S^{i-1})- H( Y_{i},S_i|Y^{i-1},S^{i-1},X^i)\nonumber \\
&\stackrel{(b)}{=} &\frac{1}{N}
\max_{Q(x^N||z^{N-1})} \sum_{i=1}^{N}H( Y_{i},S_i|Y^{i-1}, S^{i-1})-H( Y_{i},S_i|S_{i-1},X_i) \nonumber \\
&\stackrel{(c)}{\leq} &\frac{1}{N}
\max_{Q(x^N||z^{N-1})} \sum_{i=1}^{N}H( Y_{i},S_i|S_{i-1})-H( Y_{i},S_i|S_{i-1},X_i) \nonumber \\
&\stackrel{(d)}{=} &\frac{1}{N} \max_{\{Q(x_i|s_{i-1})\}}
\sum_{i=1}^{N} I( Y_{i},S_i;X_i|S_{i-1}).
\end{eqnarray}
Equality (a) follows by replacing $L_i$ with $S_i$ according to
the communication setting. Equality (b) follows from the FSC
property. Inequality (c) holds because conditioning reduces
entropy. Equality (d) holds because maximizing over the set of
causal conditioning probability $Q(x^N||z^{N-1})$ is the same as
maximizing over the set of probabilities
$\{Q(x_i|s_{i-1})\}_{i=1}^N$, as shown in the following argument.
The sum $\sum_{i=1}^{N} I( Y_{i},S_i;X_i|S_{i-1})$ is determined
uniquely by the sequence of probabilities
$\{P(y_i,s_i,x_i,s_{i-1}\}_{i=1}^N$. Let us prove by induction
that this sequence of probabilities is determined by
$\{Q(x_i|x^{i-1},y^{i-1},s^{i-1})\}_{i=1}^N$ only through
$\{Q(x_i|s_{i-1})\}_{i=1}^N$. For $i=1$ we have
\begin{eqnarray}
P(y_1,s_1,x_1,s_{0})=P(s_0)Q(x|s_0)p(y_1,s_1|x_1,s_0).
\end{eqnarray}
Since $P(s_0)$ and $P(y_1,s_1|x_1,s_0)$ are determined by the
channel properties, the input distribution to the channel can
influence only the term $Q(x|s_0)$. Now, let us assume that the
argument is true for $i-1$ and let us prove it for $i$.
\begin{eqnarray}
P(y_{i},s_i,x_i,s_{i-1})&=&P(s_{i-1})Q(x_i|s_{i-1})P(y_i,s_{i}|x_i,s_{i-1}).
\end{eqnarray}
The term $P(s_{i-1})$ is the same under both sequences of
probabilities because of the assumption that the argument holds
for $i-1$. The term $P(y_i,s_{i}|x_i,s_{i-1})$ is determined by
the channel, so the only term influenced by the input distribution
is $Q(x_i|s_{i-1})$. This proves the validity of the argument for
all $i$ and consequently, the equality (d).
%
%
%\begin{eqnarray}\label{e_cli5}
% %\lefteqn{
% \max_{Q(x^N||z^{N-1})} \sum_{i=1}^{N}H(
% Y_{i},S_i|S_{i-1})
% % =}\nonumber \\
%&=&\max_{Q(x^N||z^{N-1})} \sum_{i=1}^{N} \mathbf E \left[ \log P( Y_{i},S_i|S_{i-1}) \right] \nonumber \\
%&\stackrel{}{=} &
%\max_{Q(x^N||z^{N-1})} \sum_{i=1}^{N} \mathbf E \left[\log \sum_{x_i}P( Y_{i},x_i,S_i|S_{i-1})\right] \nonumber \\
%&\stackrel{}{=} &
%\max_{Q(x^N||z^{N-1})} \sum_{i=1}^{N} \mathbf E \left[\log \sum_{x_i}P( Y_{i},S_i|S_{i-1},x_i)P(x_i|S_{i-1})\right] \nonumber \\
%&\stackrel{(a)}{=} & \max_{Q(x^N||s^{N-1})} \sum_{i=1}^{N} \mathbf E
%\left[\log \sum_{x_i}P(
%Y_{i},S_i|S_{i-1},x_i)P(x_i|S_{i-1})\right]\nonumber \\
%&\stackrel{}{=} & \max_{\{Q(x_i|s_{i-1})\}} \sum_{i=1}^{N} H(
% Y_{i},S_i|S_{i-1})
%\end{eqnarray}
%Equality (a) holds because the only term that is influenced by the
%maximization is $P(x_i|S_{i-1}))$ and therefore it is enough to
%maximize over the set $Q(x^N||s^{N-1})$.

Inequality (\ref{e_cli4}) proves that the achievable rate, when
there is feedback and state information, cannot exceed
$\lim_{N\to\infty}\frac{1}{N} \max_{Q(x^N|s^{N-1})} \sum_{i=1}^{N}
I( Y_{i},S_i;X_i|S_{i-1})$. Now let us prove that if the state of
the channel is known at the encoder and the decoder and there is
no feedback, we can achieve this rate. For this setting we denote
the capacity as $C^{(NF)}$ and as in the case of feedback, the
capacity does not depend on the initial state and is given as
$\lim_{N\to \infty}C_N^{(NF)}$, where $C_N^{(NF)}$ satisfies
\begin{eqnarray}\label{e_cli6} C_N & \stackrel{}{=} &\frac{1}{N} \max_{Q(x^N||z^{N-1})} I(X^N
\rightarrow \{Y^N,L^N\})
 \nonumber \\
&\stackrel{(a)}{=} &\frac{1}{N}
\max_{Q(x^N||s^{N-1})} \sum_{i=1}^{N}I(X^i; Y_{i},S_{i}|Y^{i-1}, S^{i-1}) \nonumber \\
&\stackrel{}{=} &\frac{1}{N}
\max_{Q(x^N||s^{N-1})} \sum_{i=1}^{N}H( Y_{i},S_i|Y^{i-1}, S^{i-1})- H( Y_{i},S_i|Y^{i-1},S^{i-1},X^i)\nonumber \\
&\stackrel{(b)}{=} &\frac{1}{N}
\max_{Q(x^N||s^{N-1})} \sum_{i=1}^{N}H( Y_{i},S_i|Y^{i-1}, S^{i-1})-H( Y_{i},S_i|S_{i-1},X_i) \nonumber \\
&\stackrel{(c)}{\geq} &\frac{1}{N}
\max_{\{Q(x_i|s_{i-1})\}} \sum_{i=1}^{N}H( Y_{i},S_i|Y^{i-1}, S^{i-1})-H( Y_{i},S_i|S_{i-1},X_i) \nonumber \\
&\stackrel{(d)}{=} &\frac{1}{N}
\max_{\{Q(x_i|s_{i-1})\}} \sum_{i=1}^{N}H( Y_{i},S_i|S_{i-1})-H( Y_{i},S_i|S_{i-1},X_i) \nonumber \\
&\stackrel{}{=} &\frac{1}{N} \max_{\{Q(x_i|s_{i-1})\}}
\sum_{i=1}^{N} I( Y_{i},S_i;X_i|S_{i-1})\nonumber \\
&\stackrel{(e)}{\geq}& C_N^{(F)}.
\end{eqnarray}
Equality (a) follows by replacing  $L_i$ and $Z_i$ with $S_i$
according to the communication setting. Equality (b) follows from
the FSC property. Inequality (c) holds because we restrict the
range of probabilities over which the maximization is performed.
Equality (d) holds because under an input distribution
$Q(x_i|s_{i-1})$, we have the following Markov chain: $
(Y_{i},S_i)-S_{i-1}-(Y^{i-1},S^{i-2})$. Inequality (e) holds due
to (\ref{e_cli4}).

Taking the limit $N\to \infty$ on both sides of (\ref{e_cli6})
shows that $C^{(NF)}\geq C^{(F)}$. Since trivially also
$C^{(NF)}\leq C^{(F)}$ we have  $C^{(NF)}= C^{(F)}$.
\end{proof}

\begin{figure}{
 \psfrag{w0\r}[][][0.8]{$u^N$}\psfrag{d0\r}[][][0.8]{Source}
\psfrag{c0\r}[][][0.8]{encoder}
\psfrag{w1a\r}[][][0.8]{$m(u^N)$}%\in\{1...2^{NR}\}$}
\psfrag{w1\r}[][][0.8]{$\in \{1...2^{NR}\}$}
 \psfrag{d1\r}[][][0.8]{Source}
 \psfrag{e1\r}[][][0.8]{channel}
\psfrag{c1\r}[][][0.8]{encoder} \psfrag{w2\r}[][][0.8]{$x_i$}
 \psfrag{d2\r}[][][0.8]{channel}
\psfrag{w3\r}[][][0.8]{$y_i$}\psfrag{e3\r}[][][0.8]{channel}
\psfrag{d3\r}[][][0.8]{Source} \psfrag{c3\r}[][][0.8]{decoder}
\psfrag{v5a\r}[][][0.8]{$\hat
m(y^N)$}%\in\{1...2^{NR}\}$}
\psfrag{v5b\r}[][][0.8]{$\in\{1...2^{NR}\}$}
\psfrag{d4\r}[][][0.8]{Source} \psfrag{c4\r}[][][0.8]{decoder}
\psfrag{v6\r}[][][0.8]{$v^N$} \psfrag{v4\r}[][][0.8]{$y_i$}
\psfrag{v6\r}[][][0.8]{$v^N$}
\psfrag{c5\r}[][][0.8]{feedback}\psfrag{d5\r}[][][0.8]{Time-invariant}
\psfrag{v7\r}[][][0.8]{$z_{i-1}(y_{i-1})$}

 \centering
\includegraphics[width=9.5cm]{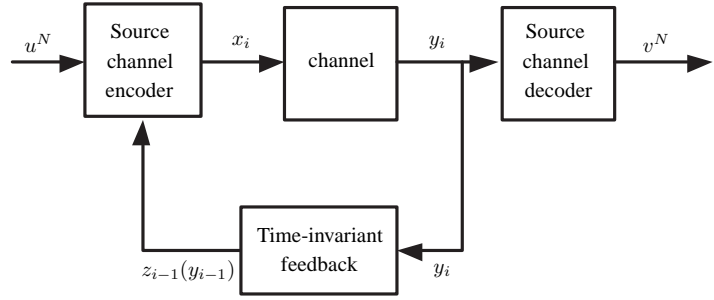}
\centering \caption{Source and channel coding, where the channel
has time-invariant feedback.} \label{f_4}
}\end{figure} %%when I save on visio I used AI 3 - and I had to restart the computer and not all visio features work with AI3

\begin{figure}{
 \psfrag{w0\r}[][][0.8]{$u^N$}\psfrag{d0\r}[][][0.8]{Source}
\psfrag{c0\r}[][][0.8]{encoder}
\psfrag{w1a\r}[][][0.8]{$m(u^N)$}%\in\{1...2^{NR}\}$}
\psfrag{w1\r}[][][0.8]{$\in \{1...2^{NR}\}$}
 \psfrag{d1\r}[][][0.8]{Channel}
\psfrag{c1\r}[][][0.8]{encoder} \psfrag{w2\r}[][][0.8]{$x_i$}
 \psfrag{d2\r}[][][0.8]{channel}
\psfrag{w3\r}[][][0.8]{$y_i$}\psfrag{d3\r}[][][0.8]{Channel}
\psfrag{c3\r}[][][0.8]{decoder} \psfrag{v5a\r}[][][0.8]{$\hat
m(y^N)$}%\in\{1...2^{NR}\}$}
\psfrag{v5b\r}[][][0.8]{$\in\{1...2^{NR}\}$}
\psfrag{d4\r}[][][0.8]{Source} \psfrag{c4\r}[][][0.8]{decoder}
\psfrag{v6\r}[][][0.8]{$v^N$} \psfrag{v4\r}[][][0.8]{$y_i$}
\psfrag{v6\r}[][][0.8]{$v^N$}
\psfrag{c5\r}[][][0.8]{feedback}\psfrag{d5\r}[][][0.8]{Time-invariant}
\psfrag{v7\r}[][][0.8]{$z_{i-1}(y_{i-1})$}

 \centering
\includegraphics[width=16cm]{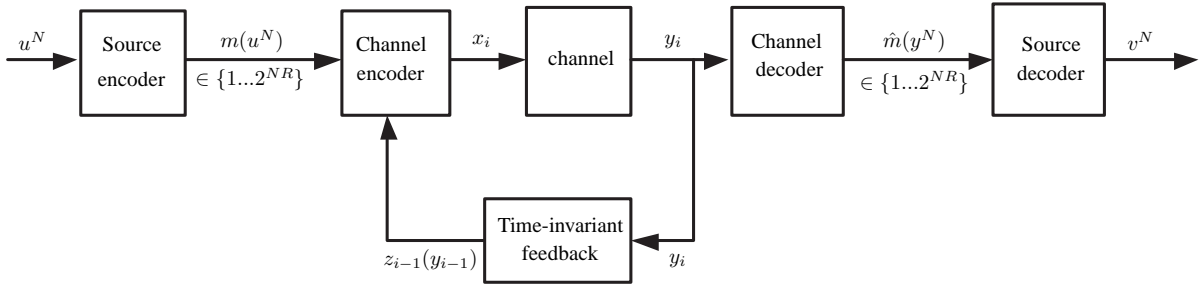}
\centering \caption{Source and channel separation.} \label{f_5}
}\end{figure} %%when I save on visio I used AI 3 - and I had to restart the computer and not all visio features work with AI3

\section{Source-channel separation \label{s_source_channel_sep}}
In this section we prove the optimality of source channel
separation for the case of an ergodic source that is transmitted
through a channel with a deterministic time-invariant feedback.
Namely, we prove that in the communication setting presented in
Fig. \ref{f_5}, the number of bits per channel use that can be
transmitted and reconstructed within a given distortion is the
same as in the communication setting of Fig \ref{f_4}.

Let us state the source-channel separation theorem as presented in
\cite[Chapter 7]{Ber71}.

\begin{theorem} Let $\epsilon>0$ and $D\geq 0$ be given. Let $R(\cdot)$ be the rate
distortion function of a discrete, stationary, ergodic source with
respect to a single letter criterion generated by a bounded
distortion measure $\rho$. Then the source output can be
reproduced with fidelity $D$ at the receiving end of any channel
if $C>R(D)$. Conversely, fidelity $D$ is unattainable at the
receiving end of any channel of capacity $C<R(D)$.

Remark: For the simplicity of the presentation we assumed one
channel use per source symbol. Our derivation below extends to the
general case where the average number of channel uses per letter
is $\frac{\tau_s}{\tau_c}$, analogously as in \cite[chapter
9]{Gallager68}.
\end{theorem}

The purpose of this section is to prove the theorem for a channel
with time-invariant feedback, as shown in Fig. \ref{f_4}, for the
cases where its capacity is given by
 \begin{equation}
 C =\lim_{N \to \infty} \frac{1}{N}  \max_{Q(x^N||z^{N-1})} I(X^N
\rightarrow Y^N ).
 \end{equation}
In the case of no feedback the proof of separation optimality is
based on data processing inequality which states that
$I(U^N;V^N)\leq I(X^N;Y^N)$ because of the Markov form
$U^N-X^N-Y^N-V^N$. However, the regular data processing inequality
does not hold for the directed information and therefore an
explicit derivation of the inequality $I(U^N;V^N)\leq I(X^N \to
Y^N)$ is needed.
\begin{proof}
The direct proof, namely that if $C>R(D)$ it is possible to
reproduce the source with fidelity $D$ is the same as for the case
without feedback\cite[Theorem 7.2.6]{Ber71}.

For the converse, namely that $R(D)$ has to be less or equal $C$,
we use the fact that for any $i$, the Markov chain
$U^N-X_i(U^N,Y^{i-1})-Y_i$ holds.
\begin{eqnarray}
NR(D) &\stackrel{(a)}{\leq} &  I(U^N;V^N) \nonumber \\
&\stackrel{(b)}{\leq} &  I(U^N;Y^N) \nonumber \\
&\stackrel{(c)}{=} & \sum_{i=1}^N I(U^N;Y_i|Y^{i-1}) \nonumber \\
&\stackrel{=}{}& \sum_{i=1}^N H(Y_i|Y^{i-1})-H(Y_i|U^N,Y^{i-1}) \nonumber \\
&\stackrel{(d)}{=}& \sum_{i=1}^N H(Y_i|Y^{i-1})-H(Y_i|U^N,Y^{i-1},X^i) \nonumber \\
&\stackrel{(e)}{=}& \sum_{i=1}^N H(Y_i|Y^{i-1})-H(Y_i|Y^{i-1},X^i) \nonumber \\
&\stackrel{=}{}& \sum_{i=1}^N I(Y_i;X^i|Y^{i-1}) \nonumber \\
&\stackrel{=}{} & I(X^N \to Y^N) \nonumber \\
&\stackrel{(f)}{\leq}  & NC.
\end{eqnarray}
Inequality (a) follows the converse for rate distortion
\cite[Theroem 7.2.5]{Ber71}. Inequality (b) follows the data
processing inequality because $U^N-Y^N-V^N$ form a Markov chain.
Equality (c) follows the chain rule. Inequality (d) follows the
fact that $X_i$ is a deterministic function of $(U^N,Y^{i-1})$.
Inequality (e) follows the Markov chain
$U^N-X_i(U^N,Y^{i-1})-Y_i$. Finally, inequality (f) follows the
converse of channel with feedback given in Theorem \ref{t_Cupper}.
\end{proof}

\section{Conclusion and future work} \label{s_conclusion}
We determined achievable rate and the capacity upper bound of FSCs
with feedback that is a deterministic function of the channel
output. The achievable rate is obtained via a random generated
coding scheme that utilizes feedback, along with a ML decoder.
%*considered the case where the feedback is a deterministic function
%of the output. We stated the achievable rate and an upper bound on
%the capacity for this communication setting.
In the case that the channel is an indecomposable FSC without ISI,
the upper bound and the achievable rate coincide and, therefore,
they are the capacity of the channel. One future direction is to
generalize the channels for which the achievable rate equals to
the upper bound on the capacity and to use this formula in order
to compute the capacity in various settings involving side
information and feedback.

By using the directed information formula for the capacity of FSCs
with feedback developed in this work, it was shown in
\cite{PermuterAllerton06} that the feedback capacity of a channel
introduced by David Blackwell in 1961 \cite{Blackwell61}, also
known as the trapdoor channel \cite{Ash65}, is the logarithm of
the golden ratio. The capacity of Blackwell's channel without
feedback is still unknown. Another future work is to find the
capacity of additional channels with time-invariant feedback.
%n


\section*{Acknowledgment}

The authors would like to thank T. Cover, G. Kramer, A. Lapidoth,
T. Moon and S. Tatikonda,  for helpful discussions, and are
indebted to Young-Han Kim for suggesting a simple proof of Lemma
\ref{l_dirB}.

\bibliographystyle{unsrt}
\bibliography{feedback2_arxive}

\begin{thebibliography}{10}

\bibitem{shannon56}
C.~E. Shannon.
\newblock The zero error capacity of a noisy channel.
\newblock {\em IRE Trans. Information Theory}, IT-2:8--19, 1956.

\bibitem{Gallager68}
R.~C. Gallager.
\newblock {\em Information Theory and Reliable Communication}.
\newblock Wiley, New York, 1968.

\bibitem{Blackwell58}
L.~Breiman D.~Blackwell and A.J. Thomasian.
\newblock Proof of shannon's transmission theorem for finite-state
  indecomposable channels.
\newblock {\em Ann. Math. Stat}, 29:1209--2220, 1958.

\bibitem{Cover89}
T.~M. Cover and S.~Pombra.
\newblock Gaussian feedback capacity.
\newblock {\em IEEE Trans. Inform. Theory}, 35(1):37--43, 1989.

\bibitem{Massey90}
J.~Massey.
\newblock Causality, feedback and dircted information.
\newblock {\em Proc. Int. Symp. Information Theory Application (ISITA-90)},
  pages 303--305, 1990.

\bibitem{Marko73}
H.~Marko.
\newblock The bidirectional communication theory- a generalization of
  information theory.
\newblock {\em IEEE Tran. on communication}, COM-21:1335--1351, 1973.

\bibitem{Tatikonda00}
S.C. Tatikonda.
\newblock Control under communication constraints.
\newblock {\em Ph.D. disertation, MIT, Cambridge, MA}, 2000.

\bibitem{Verdu94}
S.~Verd{\'u} and F.~Han.
\newblock A general formula for channel capacity.
\newblock {\em IEEE. Trans. Inform. Theory}, 40:1147--1157, 1994.

\bibitem{Yang05}
S~Tatikonda S~Yang, A~Kavcic.
\newblock Feedback capacity of finite-state machine channels.
\newblock {\em IEEE Transactions on Information Theory}, pages 799--810, 2005.

\bibitem{Chen05}
J.~Chen and T.~Berger.
\newblock The capacity of finite-state markov channels with feedback.
\newblock {\em IEEE Tran. on Information theory}, 51:780--789, 2005.

\bibitem{Weissman03}
N.~Merhav T.~Weissman.
\newblock On competitive prediction and its relation to rate-distortion theory.
\newblock {\em IEEE Trans. Inform. Theory}, 49(12):3185--3194, 2003.

\bibitem{Pradhan04}
S.S. Pradhan.
\newblock Source coding with feedforward: Gaussian sources.
\newblock In {\em Proceedings 2004 International Symposium on Information
  Theory}, page 212, 2004.

\bibitem{Pradhan04b}
R.~Venkataramanan and S.~S. Pradhan.
\newblock Source coding with feedforward: Rate-distortion function for general
  sources.
\newblock In {\em IEEE Information theory workshop (ITW)}, 2004.

\bibitem{zamir06}
R.~Zamir, Y.~Kochman, and U.~Erez.
\newblock Achieving the gaussian rate-distortion function by prediction.
\newblock Submitted for publication in ``IEEE Trans. Inform. Theory", July
  2006.

\bibitem{shamai99}
G~Caire and S~Shamai.
\newblock On the capacity of some channels with channel state information.
\newblock {\em IEEE Transactions on Information Theory,}, 45:2007--2019, 1999.

\bibitem{Kramer03}
G.~Kramer.
\newblock Capacity results for the discrete memoryless network.
\newblock {\em IEEE Trans. Inform. Theory}, 49:4--21, 2003.

\bibitem{Kramer88}
G.~Kramer.
\newblock {\em Directed Information for Channels with Feedback}.
\newblock Ph.d. dissertation, Swiss Federal Institute of Technology Zurich,
  1998.

\bibitem{Massey05}
J.~Massey.
\newblock Conservation of mutual and directed information.
\newblock {\em Proc. Int. Symp. Information Theory (ISIT-05)}, pages 157--158,
  2005.

\bibitem{Blahut87}
R.~E. Blahut.
\newblock {\em Principles and Practice of Information Theory}.
\newblock Addison-Wesley, Reading, MA, 1987.

\bibitem{Shannon61}
C.E. Shannon.
\newblock Two--way communication channels.
\newblock In {\em Proc. 4th Berkeley Symp. Math. Statist. and Prob.}, pages
  611--614. Univ. of California Press, Berkeley, 1961.

\bibitem{CovThom}
T.~Cover and J.~A. Thomas.
\newblock {\em Elements of Information Theory}.
\newblock Wiley, 1991.

\bibitem{Jelinek65}
F.~Jelinek.
\newblock Indecomposable channels with side information at the transmitter.
\newblock {\em Information and Control,}, 8:36--55, 1965.

\bibitem{Doob53}
J.~L. Doob.
\newblock {\em Stochastic Processes}.
\newblock Wiley, New York, 1953.

\bibitem{Ber71}
T.~Berger.
\newblock {\em Rate Distortion Theory: A Mathematical Basis for Data
  Compression}.
\newblock Prentice-Hall, Englewood, NJ, 1971.

\bibitem{PermuterAllerton06}
H.~H. Permuter, P.~W. Cuff, B.~Van-Roy, and T.~Weissman.
\newblock Capacity of the trapdoor channel with feedback.
\newblock {\em Submitted to the 44-th Annual Allerton Conference on
  Communication, Control and Computing}, 2006.

\bibitem{Blackwell61}
D.~Blackwell.
\newblock Information theory.
\newblock {\em Modern mathematics for the engineer: Second series}, pages
  183--193, 1961.

\bibitem{Ash65}
R.~Ash.
\newblock {\em Information Theory}.
\newblock Wiley, New York, 1965.

\end{thebibliography}

\newpage

%\begin{center}
%    {\bf APPENDIX - related research that I do not know if it is publishable!}
%  \end{center}
\appendices
\section{Proof of Lemma \ref{l_dirB}}\label{a_l_dirB}
%\begin{proof}
\begin{eqnarray}
\lefteqn{\left|I(X^N \rightarrow Y^N||Z^{N-1} )- I(X^N \rightarrow
Y^N||Z^{N-1},S)\right|}\nonumber \\
& \stackrel{(a)}{=} &\left|\sum_{i=1}^{N} I(Y_i;X^i|Y^{i-1},Z^{i-1})-I(Y_i;X^i|Y^{i-1},Z^{i-1},S)\right| \nonumber\\
& \stackrel{}{=}&\left|\sum_{i=1}^{N} H(Y_i|Y^{i-1},Z^{i-1})-H(Y_i|Y^{i-1},X^{i},Z^{i-1})-H(Y_i|Y^{i-1},Z^{i-1},S) +H(Y_i|Y^{i-1},X^{i},Z^{i-1},S)\right| \nonumber\\
& \stackrel{}{=}&\left|\sum_{i=1}^{N} H(Y_i|Y^{i-1},Z^{i-1})-H(Y_i|Y^{i-1},Z^{i-1},S)-H(Y_i|Y^{i-1},X^{i},Z^{i-1}) +H(Y_i|Y^{i-1},X^{i},Z^{i-1},S)\right| \nonumber\\
& \stackrel{}{=}&\left|\sum_{i=1}^{N} I(Y_i;S|Y^{i-1},Z^{i-1})-I(Y_i;S|Y^{i-1},X^{i},Z^{i-1})\right| \nonumber\\
& \stackrel{(b)}{\leq}&\max \left(\sum_{i=1}^{N} I(Y_i;S|Y^{i-1},Z^{i-1}),\sum_{i=1}^{N}I(Y_i;S|Y^{i-1},X^{i},Z^{i-1})\right) \nonumber\\
& \stackrel{(c)}{\leq}&\max \left(\sum_{i=1}^{N} I(Y_i,Z_i;S|Y^{i-1},Z^{i-1}),\sum_{i=1}^{N}I(Y_i,Z_i,X_{i+1};S|Y^{i-1},X^{i},Z^{i-1})\right) \nonumber\\
& \stackrel{(d)}{=}&\max \left( I(Y^N,Z^N;S),I(Y^N,Z^N,X_2^N;S)\right) \nonumber\\
& \stackrel{(e)}{\leq}&\max \left(H(S),H(S) \right) \nonumber\\
& \stackrel{}{\leq}& \log |\mathcal{S}|
\end{eqnarray}

Equality (a) is due to the definition of the directed information.
Inequality (b) holds because the magnitude of the difference
between two positive numbers is smaller than the maximum of the
numbers. Inequality (c) is due to the fact that $I(X;Y) \leq
I(X,Z;Y)$ for any random variables $X,Y,Z$.
 Equality (d) is due to the chain rule of
mutual information. Inequality (e) is due to the fact that mutual
information of two variables is smaller than the entropy of each
variable, and the last inequality holds because the cardinality of
the alphabet of $S$ is $|\mathcal{S}|$.
%\end{proof}
\QED

\section{Proof of Theorem \ref{t_MLB}}\label{a_t_MLB}
%\begin{proof}
\begin{eqnarray}\label{e_pem}
\mathbf E(P_{e,m})& = & \sum_{y^N}\sum_{x^N}
P(x^N,y^N)P[error|m,x^N,y^N]
\nonumber \\
& = & \sum_{y^N}\sum_{x^N}
Q(x^N||z^{N-1})P(y^N||x^N)P[error|m,x^N,y^N],
\end{eqnarray}
where $P[error|m,x^N,y^N]$ is the probability of decoding error
conditioned on the message $m$, the output $y^N$ and the input
$x^N$. The second equality is due to Lemma \ref{l_QP}.
%scheme $x^N$ %where each bit $x_i$ is a function of the message
%$m$ and
%the feedback $z^{i-1}$.
%The Probability is with respect to the randomness in assigning the input schemes for the other messages. \\
 Throughout the reminder of the proof we fix the message $m$. For a given tuple $(m,x^N,y^N)$ define the event $A_{m'}$, for each
 $m'\neq m$, as the event that the message $m'$ is selected in such
 a way that $P(y^N|m')>P(y^N|m)$ which, according to eq.
 (\ref{e_ml}),
 is the same as $P(y^N||x'^N)> P(y^N||x^N)$ where $x'^N$
 is a shorthand notation for $x^N(m',z^{N-1}(y^{N-1}))$ and $x^N$ is a shorthand notation for $x^N(m,z^{N-1}(y^{N-1}))$.
From the definition of $A_{m'}$ we have
\begin{eqnarray}\label{e_pAm}
P(A_{m'}|m,x^N,y^N)&=&\sum_{x'^N} Q(x'^N||z^{N-1}) \cdot {\bf I}
[P(y^N||x'^N)> P(y^N||x^N )]\nonumber \\
& \leq & \sum_{x'^N}  Q (x'^N||z^{N-1}) \left[\frac{
P(y^N||x'^N)}{P(y^N||x^N)}\right]^s; \qquad \text{any } s>0
\end{eqnarray}
where ${\bf I}(x)$ denotes the indicator function.
\begin{eqnarray}\label{e_MLB2}
P[error|m,x^N,y^N] & = & P(\bigcup_{m'\neq m} A_{m'}|m,x^N,y^N)
\nonumber \\
& \leq & \min  \left\{ \sum_{m'\neq m} P(A_{m'}|m,x^N,y^N), 1 \right\}\nonumber\\
& \leq & \left[ \sum_{m'\neq m} P(A_{m'}|m,x^N,y^N) \right] ^\rho;
\qquad \text{any } 0 \leq \rho \leq 1 \nonumber\\
& \leq & \left[ (M-1) \sum_{x'^N} Q (x'^N||z^{N-1}) \left[\frac{
P(y^N||x^N)}{P(y^N||x'^N)}\right]^s\right] ^\rho, \qquad 0 \leq
\rho \leq 1, s>0,
\end{eqnarray}
where the last inequality is due to inequality (\ref{e_pAm}). By
substituting inequality (\ref{e_MLB2}) in eq. (\ref{e_pem}) we
get:
\begin{eqnarray}
{\bf E}[P_{e,m}] \leq (M-1)^{\rho}\sum_{y^N}\left[ \sum_{x^N}
Q(x^N||z^{N-1})P(y^N||x^N)^{1-s\rho}\right]\left[ \sum_{x'^N} Q
(x'^N||z^{N-1})P(y^N||x'^N)^{s}\right]^\rho.
\end{eqnarray}
By substituting $s=1/(1+\rho)$, and recognizing that $x'$ is a
dummy variable of summation, we obtain eq. (\ref{e_MLB}) and
complete the proof.
%\end{proof}
\QED

\section{Proof of Theorem \ref{t_fn}} \label{a_t_fn}
 Theorem \ref{t_MLB} holds for
any distribution of the initial states $S_0$. In particular, it
holds for the case that $P(s_0)=\frac{1}{|\mathcal{S}|}$, namely,
the uniform distribution. By assuming a uniform distribution on
the initial state, we get that the likelihood function
satisfies%, for $m,y^N$ and $x^N=x^N(m,z^{N-1})$%
%\begin{equation}
% P(y^N||x^N) =
%\sum_{i=1} \frac{1}{A} P(y^N||x^N,s_0)
%\end{equation}
%This equality does not hold in general for any probability
%distribution $P(y^N,x^N)$ but it does hold in the case of the
%feedback channel setting.
\begin{eqnarray}
P(y^N||x^N) & \stackrel{(a)}{=}& P(y^N|m) \nonumber\\
& = &  \sum_{s_0} P(y^N,s_0|m) \nonumber\\
& \stackrel{(b)}{=}& \sum_{s_0} P(s_0) P(y^N|m,s_0)  \nonumber\\
& \stackrel{}{=}& \sum_{s_0} \frac{1}{|\mathcal{S}|} \prod_{i=1}^{N} P(y_i|y^{i-1},m,s_0)  \nonumber\\
& \stackrel{}{=}& \sum_{s_0} \frac{1}{|\mathcal{S}|} \prod_{i=1}^{N} P(y_i|y^{i-1},m,x^i,s_0)  \nonumber\\
& \stackrel{}{=}& \sum_{s_0} \frac{1}{|\mathcal{S}|} \prod_{i=1}^{N} P(y_i|y^{i-1},x^i,s_0)  \nonumber\\
& \stackrel{}{=}& \sum_{s_0} \frac{1}{|\mathcal{S}|}
P(y^N||x^N,s_0).
\end{eqnarray}
Equality (a) is shown in eq. (\ref{e_ml}) and Equality (b) holds
due to the assumption that the initial state $S_0$ and the message
$m$ are independent. Thus, assuming that $s_0$ is uniformly
distributed the bound on error probability under ML decoding given
in Theorem \ref{t_MLB} becomes
%Under the assumption that all initial states are equally likely,
%the maximum likelihood bound given in Theorem \ref{t_MLB} becomes
%\begin{equation}\label{e_MLBs0}
%\mathbf E(P_{e,m}) \leq (M-1)^{\rho} \sum_{y^N} \left\{ \sum_{x^N}
%Q(x^N||z^{N-1}) \left[  \sum_{s_0} \frac{1}{A}
%P(y^N||x^N,s_0)\right]^{\frac{1}{(1+\rho)}} \right\} ^{1+\rho} ,
%\qquad 0 \leq \rho \leq 1.
%\end{equation}
%It can be observed that for an arbitrary initial state
%distribution and for ML decoder that is $\sum_{s_0} \frac{1}{A}
%P(y^N||x^N,s_0)$ the following bound holds:
\begin{equation}\label{e_MLBs0b}
\sum_{s_0} \frac{1}{|\mathcal{S}|} \mathbf E(P_{e,m}(s_0))
 \leq (M-1)^{\rho} \sum_{y^N} \left\{ \sum_{x^N}
Q(x^N||z^{N-1}) \left[  \sum_{s_0} \frac{1}{|\mathcal{S}|}
P(y^N||x^N,s_0)\right]^{\frac{1}{(1+\rho)}} \right\} ^{1+\rho} ,
\qquad 0 \leq \rho \leq 1
\end{equation}
And therefore, for any initial state $s_0$
\begin{equation}\label{e_MLBs0c}
 \mathbf E(P_{e,m}(s_0))
 \leq |\mathcal{S}| (M-1)^{\rho} \sum_{y^N} \left\{ \sum_{x^N}
Q(x^N||z^{N-1}) \left[  \sum_{s_0} \frac{1}{|\mathcal{S}|}
P(y^N||x^N,s_0)\right]^{\frac{1}{(1+\rho)}} \right\} ^{1+\rho} ,
\qquad 0 \leq \rho \leq 1
\end{equation}
Since $m$ was arbitrary, we obtain a fortiori
\begin{equation}\label{e_MLBs0c}
 \mathbf E(P_{e}(s_0))
 \leq |\mathcal{S}| (M-1)^{\rho} \sum_{y^N} \left\{ \sum_{x^N}
Q(x^N||z^{N-1}) \left[  \sum_{s_0} \frac{1}{|\mathcal{S}|}
P(y^N||x^N,s_0)\right]^{\frac{1}{(1+\rho)}} \right\} ^{1+\rho} ,
\qquad 0 \leq \rho \leq 1
\end{equation}
where $P_{e}(s_0)$ is the probability of error over all messages
given that the initial state is $s_0$ and the expectation is w.r.t
the random generation of the code. It is possible to construct a
code for $2M$ messages that this inequality holds for the average
and then to pick the best $M$ messages such that the bound holds
for each message within a factor of 4. I.e., we get that for every
$1\leq m \leq M$,
\begin{equation}\label{e_MLBs0d}
 P_{e,m}(s_0)
 \leq 4 |\mathcal{S}| (M-1)^{\rho} \sum_{y^N} \left\{ \sum_{x^N}
Q(x^N||z^{N-1}) \left[  \sum_{s_0} \frac{1}{|\mathcal{S}|}
P(y^N||x^N,s_0)\right]^{\frac{1}{(1+\rho)}} \right\} ^{1+\rho} ,
\qquad 0 \leq \rho \leq 1
\end{equation}
By using the inequality $(\sum_i a_i)^r\leq \sum_i (a_i)^r$ for
$0\leq r \leq 1$ we can move the sum over $s_0$, yielding
\begin{equation}\label{e_MLBs0d}
 P_{e,m}(s_0)
 \leq 4 |\mathcal{S}| (M-1)^{\rho} \sum_{y^N} \left\{ \sum_{s_0} \sum_{x^N}
Q(x^N||z^{N-1}) \left[   \frac{1}{|\mathcal{S}|}
P(y^N||x^N,s_0)\right]^{\frac{1}{(1+\rho)}} \right\} ^{1+\rho} ,
\qquad 0 \leq \rho \leq 1
\end{equation}
Furthermore, we can move the sum over $s_0$ once again by
rearranging the sum and then using the Jensen's inequality:
\begin{eqnarray}\label{e_MLBs0e}
 P_{e,m}(s_0)
& \stackrel{(a)}{\leq}&  4|\mathcal{S}| (M-1)^{\rho} \sum_{y^N}
\left\{|\mathcal{S}|^\frac{\rho}{\rho+1} \sum_{s_0}
\frac{1}{|\mathcal{S}|} \sum_{x^N}
Q(x^N||z^{N-1}) \left[    P(y^N||x^N,s_0)\right]^{\frac{1}{(1+\rho)}} \right\} ^{1+\rho}  \nonumber\\
%----------
& \stackrel{(b)}{=}& 4 |\mathcal{S}| (M-1)^{\rho}
|\mathcal{S}|^\rho \sum_{y^N} \left\{
\sum_{s_0} \frac{1}{|\mathcal{S}|} \sum_{x^N} Q(x^N||z^{N-1}) \left[ P(y^N||x^N,s_0)\right]^{\frac{1}{(1+\rho)}} \right\} ^{1+\rho}  \nonumber\\
%----------
& \stackrel{(c)}{\leq}&  4|\mathcal{S}| (M-1)^{\rho}
|\mathcal{S}|^\rho \sum_{y^N} \sum_{s_0}
\frac{1}{|\mathcal{S}|} \left\{ \sum_{x^N} Q(x^N||z^{N-1}) \left[ P(y^N||x^N,s_0)\right]^{\frac{1}{(1+\rho)}} \right\} ^{1+\rho}  \nonumber\\
%----------
& \stackrel{}{\leq}&  4 (M-1)^{\rho} |\mathcal{S}|^\rho \sum_{s_0}
\sum_{y^N}
 \left\{ \sum_{x^N} Q(x^N||z^{N-1}) \left[ P(y^N||x^N,s_0)\right]^{\frac{1}{(1+\rho)}} \right\} ^{1+\rho}  \nonumber\\
%----------
& \stackrel{(d)}{\leq}&  4 |\mathcal{S}|(M-1)^{\rho}
|\mathcal{S}|^\rho \max_{s_0} \sum_{y^N}
 \left\{ \sum_{x^N} Q(x^N||z^{N-1}) \left[ P(y^N||x^N,s_0)\right]^{\frac{1}{(1+\rho)}} \right\} ^{1+\rho}
%----------
\end{eqnarray}
Inequalities (a) and (b) are achieved by moving the the term
$\frac{1}{|\mathcal{S}|}$ outside the sums. Inequality (c) is
achieved by applying Jensen's inequality $(\sum_i P_ia_i)^r\leq
\sum_i P_i(a_i)^r$.
%Under the assumption of random coding we can minimize the bound by
%changing the probability assignment function $Q(x^N||z^{N-1})$.
Inequality (d) holds because the number of elements multiplied by
the maximum element is larger than the sum of elements. Because
the inequality holds for all $Q(x^N||z^{N-1})$,
\begin{eqnarray}\label{e_MLBs0f}
 P_{e,m}(s_0) & \stackrel{}{\leq}&   4|\mathcal{S}|(M-1)^{\rho} |\mathcal{S}|^\rho \min_{Q(x^N||z^{N-1})}
\max_{s_0} \sum_{y^N}
 \left\{ \sum_{x^N} Q(x^N||z^{N-1}) \left[ P(y^N||x^N,s_0)\right]^{\frac{1}{(1+\rho)}} \right\} ^{1+\rho}
\end{eqnarray}
By substituting $M=2^{NR}$ and eq. (\ref{eq_fn}) and (\ref{eq_E})
into (\ref{e_MLBs0f}), we prove the theorem.
%\end{proof}
\QED

\section{Proof of Lemma \ref{l_fn}}\label{a_l_fn} Let us divide the input $x^N$
into two sets ${\bf x_1} = x_1^n$ and ${\bf x_2}= x_{n+1}^N$.
Similarly, let us divide the output $y^N$ into two sets ${\bf y_1}
= y_1^n$ and ${\bf y_2}= y_{n+1}^N$ and the feedback $z^N$ into
${\bf z_1} = z_1^{n-1}$ and ${\bf z_2}= z_{n+1}^{N-1}$. Let
$Q_n({\bf x_1}||{\bf z_1})=\prod_{i=1}^{n} P(x_i|x^i,z^{i-1})$ and
$ Q_l({\bf x_2}||{\bf z_2})
=\prod_{i=1}^{l}P(x_{n+i}|x_{n+1}^{n+i},z_{n+1}^{n+i-1})$ be the
probability assignments that achieve the maxima $F_n(\rho)$ and
$F_l(\rho)$, respectively. Let us consider the probability
assignment $ Q(x^N||z^{N-1})=Q_n({\bf x_1}||{\bf z_1}) Q_l({\bf
x_2}||{\bf z_2})$. Then
\begin{equation}\label{e_FN_ine}
F_N \geq \frac{\rho \log |\mathcal{S}|}{N}+
E_{o,N}(\rho,Q(x^N||z^{N-1},s_0')
\end{equation}
where $s_0'$ is the state that minimizes $E_{o,N}(\rho,Q
(x^N||z^{N-1}),s_0')$.

Now,
\begin{eqnarray} %\label{e_d}
P(y^N||x^N,s_0') & \stackrel{(a)}{=}& P(y^N|m,s_0') \nonumber \\
 & \stackrel{}{=}& \sum_{s_n} P(y^N,s_n|m,s_0') \nonumber \\
 & \stackrel{}{=}& \sum_{s_n} P({\bf y_1},s_n|m,s_0') P({\bf y_2}|m,s_n,{\bf y_1},s_0') \nonumber \\
 & \stackrel{}{=}& \sum_{s_n} P({\bf y_1},s_n|m,s_0')  P({\bf y_2}||{\bf x_2},s_n) \end{eqnarray}
Equality (a) can be proved in the same way as eq. (\ref{e_ml}) was
proved. The term $P({\bf y_1},s_n|m,s_0')$ can be also expressed
in terms of $\bf {y_1, x_1}$ in the following way:

\begin{eqnarray} \label{e_y1}
P({\bf y_1},s_n|m,s_0') & \stackrel{}{=}& P(s_n|m,{\bf y_1},s_0')P({\bf y_1}|m,s_0') \nonumber \\
 & \stackrel{}{=}& P(s_n|{\bf x_1},{\bf y_1},s_0') P({\bf y_1}||{\bf x_1},s_0)
\end{eqnarray}

hence we obtain:
\begin{equation}\label{e_yN}
P(y^N||x^N,s_0')=\sum_{s_n}  P({\bf y_2}||{\bf x_2},s_n)
P(s_n|{\bf x_1},{\bf y_1},s_0') P({\bf y_1}||{\bf x_1},s_0)
\end{equation}
Consequently,
\begin{eqnarray}\label{e_fn_sub}
\lefteqn{2^{[-NF_N(\rho)]}  } \nonumber \\
&\stackrel{(a)}{\leq} & |\mathcal{S}|^{\rho} \sum_{y^N}\left[
\sum_{x^N}Q (x^N||z^{N-1})P(y^N||x^N,s_0')^{1/(1+\rho)}
\right]^{1+\rho} \nonumber \\
%-------------
& \stackrel{(b)}{=}&  |\mathcal{S}|^{\rho} \sum_{{\bf y_1}{\bf
y_2}}\left\{ \sum_{{\bf x_1}{\bf x_2}} Q({\bf x_1}||{\bf z_1})
Q({\bf x_2}||{\bf z_2}) \left[ \sum_{s_n}  P({\bf y_2}||{\bf
x_2},s_n) P(s_n|{\bf x_1},{\bf y_1},s_0') P({\bf y_1}||{\bf
x_1},s_0)
  \right]^{1/(1+\rho)}
\right\}^{1+\rho} \nonumber \\
%-------------
& \stackrel{(c)}{\leq}&  |\mathcal{S}|^{2\rho} \sum_{s_n}
\sum_{{\bf y_1}{\bf y_2}}\left\{ \sum_{{\bf x_1}{\bf x_2}}Q({\bf
x_1}||{\bf z_1}) Q({\bf x_2}||{\bf z_2}) \left[   P({\bf
y_2}||{\bf x_2},s_n) P(s_n|{\bf x_1},{\bf y_1},s_0') P({\bf
y_1}||{\bf x_1},s_0)
  \right]^{1/(1+\rho)}
\right\}^{1+\rho} \nonumber \\
%-------------
& \stackrel{}{\leq}&  |\mathcal{S}|^{2\rho} \sum_{s_n} \sum_{\bf
y_1 y_2}\left[ \sum_{\bf x_1}Q({\bf x_1}||{\bf z_1}) [P(s_n|{\bf
x_1},{\bf y_1},s_0') P({\bf y_1}||{\bf x_1},s_0)]
  ^{1/(1+\rho)} \right]^{1+\rho}
  %--
  %\sum_{\bf y_2}
  \left[ \sum_{\bf x_2}Q({\bf x_2}||{\bf z_2})   P({\bf
y_2}||{\bf x_2},s_n) ^{1/(1+\rho)} \right]^{1+\rho}
   \nonumber \\
%-------------
& \stackrel{}{\leq}&  |\mathcal{S}|^{2\rho} \sum_{s_n} \sum_{\bf
y_1}\left[ \sum_{\bf x_1}Q({\bf x_1}||{\bf z_1}) [P(s_n|{\bf
x_1},{\bf y_1},s_0') P({\bf y_1}||{\bf x_1},s_0)]
  ^{1/(1+\rho)} \right]^{1+\rho}
  %--
2^{[-lF_l(\rho)]}\nonumber \\
%-------------
& \stackrel{(d)}{\leq} &  |\mathcal{S}|^{\rho}  \sum_{\bf y_1}
\left\{ \sum_{\bf x_1}Q({\bf x_1}||{\bf z_1}) \left[\sum_{s_n}
P(s_n|{\bf x_1},{\bf y_1},s_0') P({\bf y_1}||{\bf x_1},s_0)
  \right]^{1/(1+\rho)} \right\}^{1+\rho}
2^{[-lF_l(\rho)]} \nonumber \\
%-------------
& \stackrel{}{\leq} & 2^{[-nF_n(\rho)-lF_l(\rho)]}
\end{eqnarray}
Inequality (a) is due to inequality \ref{e_FN_ine}. Equality (b)
is due to eq. (\ref{e_yN}). Inequality (c) holds because of the
same reason as given in eq. (\ref{e_MLBs0d}), namely $(\sum_i
a_i)^r\leq \sum_i (a_i)^r$. Inequality (d) is due to Minkowski's
inequality $\left[\sum_jP_j \left( \sum_k a_{jk}\right)
^{1/r}\right] ^r \geq \sum_k \left( \sum_j P_ja_{jk} ^{1/r}\right)
^r$ for $r>1$. \QED

\end{document}